\documentclass{aa}  
\usepackage{graphicx}
\usepackage{txfonts}
\usepackage{hyperref}
\usepackage[dvipsnames]{xcolor}
\begin{document}

   \title{The bulk metal content of WASP-80 b from joint interior–atmosphere retrievals}

   \subtitle{Breaking degeneracies and exploring biases with panchromatic spectra}

   \author{L. Acu\~{n}a-Aguirre\inst{1}
          \and
          L. Kreidberg\inst{1}
          \and 
          P. Mollière\inst{1}
          \and
          N. Bachmann\inst{1}
          }

   \institute{Max Planck Institut für Astronomie, Königstuhl 17, 69117 Heidelberg, Germany\\
              \email{acuna@mpia.de}
       }

   %\date{Received September 15, 1996; accepted March 16, 1997}

   \date{Received 21 July 2025; accepted X}

% \abstract{}{}{}{}{} 
% 5 {} token are mandatory

% often called "super-puffs", present a challenge to planet formation theories due to the discrepancy between their extremely low bulk metal content and metal-rich envelopes.
  \abstract{
   The atmospheres of warm gas giants can be readily characterized through transmission and emission spectroscopy. WASP-80 b is one such exoplanet, with an unusually low density that is in tension with the metal-rich composition expected for a planet of this mass. 
   %WASP-80 b is one such exoplanet, \textbf{with half Jupiter’s mass and a Jupiter-like radius, making its low density difficult to explain with a super-solar atmosphere and a core mass typical of the sub-Saturn regime.}
   %Warm gas giants with low densities present a unique opportunity to constrain their bulk metal mass fractions and atmospheric compositions, offering insights into their formation pathways. WASP-80 b is one such exoplanet, with panchromatic spectra from JWST and HST in both transmission (0.5$-$4 $\mu$m) and emission (1$-$12 $\mu$m), along with precise mass and age measurements.
   We aim to derive precise constraints on WASP-80 b’s bulk metal mass fraction, atmospheric composition, and thermal structure. 
   %Our goal is to derive precise constraints on the planet's bulk metal mass fraction, atmospheric composition and thermal structure.
  We conducted a suite of retrievals using three approaches: traditional interior-only, atmosphere-only, and joint interior–atmosphere retrievals. We coupled the open-source models GASTLI and petitRADTRANS, which describe planetary structure and thermal evolution, and atmospheric chemistry and clouds, respectively. Our retrievals combine mass and age with panchromatic spectra from JWST and HST in both transmission (0.5–4 $\mu$m) and emission (1–12 $\mu$m) as observational constraints.
   %We conducted a suite of retrievals using three approaches: traditional interior-only, atmosphere-only, and joint interior-atmosphere retrievals. For the joint retrievals, we coupled the open-source models GASTLI and petitRADTRANS, which account for planetary structure and thermal evolution, and  atmospheric chemistry and clouds, respectively. Our nested sampling retrievals incorporate transmission and emission spectra, together with mass and age, as observational constraints.
   %\textcolor{red}{Our fiducial, most data-complete joint retrieval for WASP-80 b yields an atmospheric metallicity of M/H = $2.75^{+0.88}_{-0.56} \times$ solar, a sub-solar C/O = $0.12^{+0.03}_{-0.02}$ -- indicative of accretion of water-rich material -- and high altitude clouds. We derive a precise bulk metal mass fraction of $Z_{\rm planet} = 0.12 \pm 0.02$, and a core mass of $M_{\rm core} = 3.49^{+3.49}_{-1.59} \ M_{\oplus}$, consistent with formation via core accretion.   
   We identify two fiducial scenarios. In the first, WASP-80 b has an internal temperature consistent with its age in the absence of external heating sources, and its atmosphere is in chemical equilibrium, with an atmospheric metallicity M/H = 2.75$^{+0.88}_{-0.56}\times$ solar, a bulk metal mass fraction $Z_{\rm planet}=0.12\pm0.02$, and a core mass $M_{\rm core} = 3.49^{+3.49}_{-1.59} \ M_{\oplus}$. In the second scenario, WASP-80 b may be inflated by an additional heat source -- possibly induced by magnetic fields -- with an atmospheric metallicity M/H = 10.00$^{+8.20}_{-4.75}\times$ solar, $Z_{\rm planet}=0.28\pm0.11$, and $M_{\rm core}=31.8^{+21.3}_{-17.5} \ M_{\oplus}$. The super-solar M/H and sub-solar C/O ratios in both scenarios suggest late pebble or planetesimal accretion, while additional heating is required to reconcile the data with the more massive core predicted by the core accretion paradigm. In general, joint retrievals are inherently affected by a degeneracy between atmospheric chemistry and internal structure. Together with flexible cloud treatment and an unweighted likelihood, this leads to larger uncertainties in bulk and atmospheric compositions than previously claimed. 
   %JWST and Ariel will enable this approach at a population-level to link planet formation to bulk and atmospheric composition.
   %We demonstrate that joint retrievals resolve discrepancies between traditional atmospheric retrievals and bulk density constraints,} and improve the precision of bulk metal mass fraction estimates by a factor of two when combined with precise stellar ages. JWST and upcoming missions such as PLATO and Ariel will enable population-level studies that link planet formation processes to bulk and atmospheric compositions.
   }

%Multi-geometry observations with JWST and Ariel

   %Wide wavelength emission spectra (1-12 $\mu$m) combined with transmission spectra in joint retrievals reduce bulk metal mass fraction uncertainties from 0.06 wt\% to 0.02 wt\% compared to traditional interior-only retrievals by breaking the degeneracy between temperature at low pressures ($< 0.1$ bar) and envelope composition. This precision can be improved to 0.01 wt\% if age (16\% precision) is included in the joint retrieval.

   \keywords{planets and satellites: atmospheres – planets and satellites: composition – planets and satellites: gaseous planets – planets and satellites: interiors – planets and satellites: physical evolution}

   \maketitle
%
%-------------------------------------------------------------------

% old abstract: Interior-only retrievals rule out atmospheric metallicities above $\sim$10$\times$ solar, conflicting with values inferred from free-chemistry atmospheric retrievals. We find that cloud treatments significantly impact atmospheric composition estimates. Joint retrievals using all the available data yield sub-solar atmospheric metallicities and constrain the bulk metal mass fraction to 5–10 wt\%. Including age as a constraint reduces degeneracies by limiting internal temperature, and improves the precision of the bulk metal content by twofold. Emission spectra without transmission and age data do not provide informative constraints on deep interior properties due to degeneracies in thermal structure.

% Joint interior-atmosphere retrievals resolve tensions between atmospheric and interior models for WASP-80 b, demonstrating their importance in accurately characterizing exoplanet compositions. This method enable tighter constraints on bulk composition, crucial for distinguishing formation pathways of gas giants.

\section{Introduction}

% (1) why we care about the the bulk composition of extrasolar gas giants, and its connection to planet formation; 

Extrasolar gas giants are composed of two main building blocks: hydrogen and helium (H/He) and metals, which include refractory material (rocks) and ices (e.g. H$_{2}$O, CH$_{4}$, NH$_{3}$). Depending on the planet formation mechanism of gas giants, their total metal and H/He mass fractions may vary widely. Two main formation mechanisms are accepted: core accretion and gravitational instability. In the core accretion scenario, gas giants undergo a first stage where metal-rich material is accreted, leading to the formation of a core. This can be further enriched in more metals by accretion of pebbles and planetesimals, until the core reaches a critical mass. Then runaway accretion of a gas-rich (H/He) envelope starts. The more massive the planet is, the more gas it binds gravitationally \citep{HM21,Helled23_sat}. Thus, core accretion produces massive planets that are metal-poor and dominated by H/He, or Neptune-mass planets that are rich in rocks and ices. The accretion of metal-rich solids and H/He gas occurs within the same time scale in gravitational instability. Traditional gravitational instability models suggest that gas giants formed via this mechanism would have similar bulk compositions to their protoplanetary disk and host star \citep{Helled08,Boley11}. However, if the mass of the disk is not massive enough, or rings and migration play an important role in disk evolution, gravitational instability can form planet cores as massive as in core accretion \citep{Jiang_Ormel23,Rice25}. This is why determining the bulk composition and atmospheric composition of gas giants simultaneously is key to understand their formation and evolution pathways \citep{Thorngren16,MH25}.

% (2) what are the challenges in determining the bulk metal mass fraction of extrasolar gas giants, what is currently the limiting factor and the typical uncertainties in interior retrievals; 

The bulk amount of metals can only be estimated by comparing (or retrieving) the observed mass and radius to interior structure calculations. Interior models of gas giants typically assume that the planet is stratified in a metal-rich core and an envelope dominated by H/He \citep{Burrows04,fortney07,Baraffe08,leconte12,Nettelmann13,Thorngren16,MH21,Miguel22,Baumeister23}. Age is the third observable variable that constrains the internal composition of gas giants, as these contract due to thermal cooling given their Kelvin-Helmholtz timescale. In addition, the metallicity probed in the upper atmosphere by transmission spectroscopy can be used to narrow the envelope metal mass fraction \citep{Bloot23,Acuna24_gastli_science_paper} - assuming that the composition is constant along the radius in the envelope. In retrievals, the precision with which the core mass and the total planet metal mass fraction are obtained depends on the precision of these four observable parameters. For gas-rich planets, an increase in radius precision produces greater improvements in bulk metal mass fraction estimates than mass \citep{Otegi20}. In the case of gaseous giants, the precision of the age is particularly important for young planets ($\leq 1$ Gyr) because of how rapidly their radius changes between 100 Myrs and 1 Gyr, and the degeneracy between luminosity (or internal temperature) and bulk metal mass fraction \citep{Muller23}.

% (3) challenges: degeneracies and tensions between bulk density and atmospheric composition 

% (3.1) new observables: connection between the interior and the atmosphere, particularly the observable parameters of the atmosphere - mention the case of WASP-107 b, the internal temperature and disequilibrium chemistry; 

The characterization of the bulk composition of extrasolar gas giants faces the challenges of the aforementioned degeneracies, as well as tensions between different observable parameters and techniques. Degeneracies can be solved by improving the precision of the age measurement, which depends on the spectral type of the star and their properties. Additional observables can also reduce degeneracies. For example, a lower limit on the internal temperature was inferred from disequilibrium chemistry for WASP-107 b \citep{Sing_24_WASP107b,Welbanks_WASP107b}, a warm gas giant observed in transmission spectroscopy with JWST. This confirmed that WASP-107 b has a metal-rich core, and obtain a precise core mass estimate. Furthermore, the Love number, which quantifies the gravity field and therefore the internal density distribution \citep{love1911}, is sensitive to the mass of the core, enabling us to constrain it in retrievals \citep{Kramm11,Kramm12,Dijk25}. This parameter has been measured for five extrasolar hot Jupiters \citep{Buhler16,Hardy17,csizmadia19,hellard20,Barros22,bernabo24}. The number of exoplanets whose internal temperature or Love number are available is limited given the required observational criteria \citep{Akinsanmi,Mukherjee25}. 

% (3.2) tensions: Mention the particular case of puffy gas giants, and the tensions between bulk density and inferred atmospheric composition, such as WASP-77Ab (August et al., Karalis et al.); 

% (4) Mention briefly Wilkinson et al. approach for WASP-39 b (a hot Jupiter), and finally, present the science case of WASP-80 b. Don't forget to define somewhere what 'joint retrieval' means when you talk about Wilkinson et al. paper.

% \citep{Pinhas19}
% HAT-P-12 b:

% \citep{Tsiaras18}
% HAT-P-18 b
% WASP-29 b
% WASP-69 b
% WASP-67 b

Tensions can also exist between the bulk density and the inferred atmospheric metallicity. This is the case of sub-Saturn giants with warm equilibrium temperatures ($T_{\rm eq} \leq 1000$ K) whose radii are $\sim$1 $R_{\rm Jup}$, such as HAT-P-12 b \citep{Hartman09}, HAT-P-18 b \citep{hatp12b}, WASP-29 b \citep{hellier10}, WASP-69 b \citep{anderson14},  WASP-67 b \citep{hellier12}, WASP-39 b \citep{Faedi11} and WASP-80 b \citep{Triaud13}. These exoplanets are easily characterized via transmission spectroscopy given their large radii \citep{Tsiaras18,Pinhas19}. For example, JWST's panchromatic transmission spectrum of WASP-39 b indicate that its atmospheric metallicity is bimodal between solar and $\sim 100 \times$ solar \citep{Ahrer_wasp39b,alderson_wasp39b,rustamkulov_wasp39b,feinstein_wasp39b}. To explain simultaneously its low density and high atmospheric metallicity, a core-less internal structure is not sufficient because its observed radius is higher than that of a pure envelope exoplanet with a 100 $\times$ solar composition \citep{fortney07}.

%\cite{wakeford18} estimated the atmospheric composition of WASP-39 b to be $\sim 150 \times$ solar based on retrievals of its Hubble Space Telescope (HST) transmission spectrum (0.8-1.7 $\mu$m). To explain simultaneously its low density and high atmospheric metallicity, a core-less internal structure is not sufficient because its observed radius is higher than that of a pure envelope exoplanet with a 150 $\times$ solar composition \citep{fortney07}. JWST's panchromatic transmission spectrum of WASP-39 b later confirmed that this estimate could be bimodal between solar and $\sim 100 \times$ solar \citep{Ahrer_wasp39b,alderson_wasp39b,rustamkulov_wasp39b,feinstein_wasp39b}. 

Reconciling apparent discrepancies between interior and atmospheric analyses will require a combination of additional data and more advanced modeling. In general, these discrepancies could have multiple causes, including biases in the atmospheric retrievals \citep{Bezard_K218b,fisher24_biases}, or unconstrained internal heating mechanisms that inflate the planet's radius \citep{Thorngren18}. Complementary observables to the exoplanet spectra could help clarify the planet's composition and thermal properties, but obtaining such parameters is difficult. For instance, measuring the Love numbers of warm gas giants is challenging because they are further away from their star than their hot Jupiter counterparts \citep{Hellard19_k2}. Similarly, access to internal temperature constraints can be limited by clouds or equilibrium chemistry. Thus, we require a new modeling and data analysis approach to investigate the source of the differences in interior and atmospheric models. This tension could be mitigated by combining both retrievals into a joint interior-atmosphere retrieval. A joint interior-atmosphere retrieval analysis uses coupled interior-atmospheric calculations as the forward models, and integrates the transmission or emission spectra with the observables parameters associated with the bulk density in the likelihood function of the retrieval. Such a retrieval has been applied by \cite{Wilkinson24} to the mass and a transmission spectrum dataset of WASP-39 b obtained by JWST's instrument NIRSpec-G395H. Nonetheless, their work presents two caveats: first, it did not include the effect of clouds in the transmission spectrum, and secondly, it did not consider emission and transmission spectra simultaneously to further constrain degeneracies.

The aim of this work is to use joint interior-atmosphere retrievals to test whether this framework can obtain more precise bulk composition estimates than the traditional interior structure retrievals by breaking degeneracies. To do this, we explore a suite of interior, atmosphere and joint interior-atmosphere retrievals. WASP-39 b and WASP-80 b are the two super-puff warm gas giants with the widest wavelength coverage in their spectra. Of these two, WASP-80 b is the only one with emission spectra available, thus we subsequently select it as our science case. In Sect. \ref{sec:wasp80b_data} we compile the observational data for WASP-80 b available for our retrievals. In Sect. \ref{sec:forward} we introduce the forward models for the three types of retrievals (interior-only, atmosphere-only and joint), while in Sect. \ref{sec:bayes} we describe our Bayesian framework, including the priors and the calculation of the likelihood function. We present the results of our suite of retrievals in Sect. \ref{sec:transmission_only} to \ref{sec:joint_retrievals}. We discuss the benefit of joint retrievals in contrast to previous work, and implications for the interior structure and formation of WASP-80 b in Sect. \ref{sec:discussion}. Finally, our conclusions can be found in Sect. \ref{sec:conclusions}.

\section{WASP-80 b data} \label{sec:wasp80b_data}

% This section is to briefly present all the data that is available for WASP-80 b: mass, radius, age and transmission and emission data sets from JWST and HST. Include a summary table with the values and  citations for reference.

WASP-80 b is a warm gas giant ($T_{\rm eq}$ = 825 K) with an intermediate mass between Jupiter and Saturn. Its radius is similar to that of Jupiter, but its mass is approximately half Jupiter's value (see Table \ref{tab:wasp80b_all_obs_data_summary}), constituting a puffy gas giant. It orbits a low-mass, late-type star -- late K or early M -- with an effective temperature of 4145 K at an orbital period of $\sim$3 days. Its mass has been measured via the radial velocity method and its radius has been characterized with transit photometry, with  consistent results across different ground-based data analyses \citep{Triaud13,Mancini14,Triaud15}. Table \ref{tab:wasp80b_all_obs_data_summary} shows the values we adopt and their references in our study for WASP-80 b's mass and radius.

In addition, WASP-80's age was estimated via gyrochronology to be $\sim$100 Myr. It also presents chromospheric activity, which is expected in stars at young ages \citep{Triaud13}. \cite{Gallet20} discusses how gyrochronology may yield a biased age for young stars ($<$ 100 Myr) because they may have not yet converged onto a well-behaved rotation–age sequence. In addition, they re-calculate the age of several stars via gyrochonology, including WASP-80, while accounting for star-planet tidal interactions. These interactions can spin up the stellar rotational period, biasing gyrochonology estimates towards younger ages. Thus, in this work we adopt the age estimate for WASP-80 from \cite{Gallet20}, which is 1.352$\pm$0.222 Gyr. 

WASP-80 b's short period and high planet-to-star radius ratio make it an excellent target for atmospheric characterization both in transmission and emission spectroscopy. It has been observed in both geometries by Spitzer photometry \citep{Triaud15,Wong22}. Furthermore, two low-resolution transmission datasets were obtained with the Hubble Space Telescope (HST) using the STIS and WFC3 instruments. These observations revealed absorption features in the 1.4 $\mu$m bandpass, attributed to H$_{2}$O and/or CH$_{4}$. The STIS spectrum showed a slope at optical wavelengths, likely caused by aerosols or stellar contamination. Using atmospheric retrievals on these data, \cite{Wong22} constrained the overall atmospheric metallicity to $[$M/H$]$ = $\sim$30 - 100 $\times$ solar, suggesting that the envelope is significantly enriched in metals. Similarly, WASP-80 b has been characterized by JWST with the NIRCam instrument both in transmission and emission. The analyses of the NIRCam datasets have shown a sub-solar C/O ratio and a $\sim$5 $\times$ solar metallicity. In addition, CH$_{4}$ was detected at 6$\sigma$ \citep{Bell23}. This detection confirmed earlier tentative evidence from ground-based observations \citep{Carleo22}. Furthermore, \cite{Wiser25} detect H$_{2}$O, CH$_{4}$, CO and CO$_{2}$ at high confidence by analyzing WASP-80 b's panchromatic emission spectrum from the JWST data sets. Finally, transmission spectroscopy observations have suggested the presence of condensates in the atmosphere of WASP-80 b. The composition of these condensates can be constrained through eclipse observations at short wavelengths ($\lambda < 3 \ \mu m$), which are covered by HST's WFC3 and JWST's NIRISS instruments. Such observations with these two instruments have revealed that the composition of the condensates may include Cr, Na$_{2}$S, KCl and ZnS, which are expected in warm ($T_{\rm eq}$ < 1000 K) atmospheres \citep{Jacobs23}. Furthermore, a moderately high Bond albedo was estimated for WASP-80 b, with a 1$\sigma$ interval of $A_{\rm B}$ = 0.15 - 0.40 \citep{Morel25}.

Table \ref{tab:wasp80b_all_obs_data_summary} summarizes the spectrum datasets we re-analyse in our suite of retrievals with their respective wavelength ranges, references and instruments. We discard the photometric datasets, as they cover similar wavelength ranges to the HST and JWST low-resolution spectra. We do not include the emission dataset from \cite{Morel25} in our retrievals due to the computational cost of reflected light calculations. Nonetheless, we discuss the effect of including reflected light data in our framework in the inference of the interior composition in Sect. \ref{sec:disc_previous}.

\begin{table}[]
\caption{\label{tab:wasp80b_all_obs_data_summary} All observational data used in our suite of retrievals of WASP-80 b}
\resizebox{\columnwidth}{!}{%
\begin{tabular}{lcc}
\hline
\textbf{Parameters} & \textbf{WASP-80} & \textbf{References} \\ \hline \hline
Stellar radius [$R_{\odot}$] & 0.571 & 1 \\
Stellar mass [$M_{\odot}$] & 0.57 & 1 \\
Effective temperature [K] & 4145 & 1 \\
 $\rm [Fe/H]$ & $-0.13^{+0.15}_{-0.17}$ & 1 \\
Age [Gyr] & 1.352 $\pm$ 0.222 & 2 \\ \hline 
\textbf{Parameters} & \textbf{WASP-80 b} & \textbf{References} \\ \hline \hline
Planet radius [$R_{\rm Jup}$] & 0.952$^{+0.026}_{-0.027}$ & 1 \\
Planet mass [$M_{\rm Jup}$] & 0.554$^{+0.030}_{-0.039}$ & 1 \\
Equilibrium temperature [K] & 825 & 3 \\ 
Semi-major axis [AU] & 0.03479 & 3 \\ \hline
\textbf{Datasets \& geometry} & \textbf{Wavelength [$\mu$m]} & \textbf{References} \\ \hline \hline
HST/STIS - Transmission & 0.4 - 1.0 & 4 \\
HST/WFC3 - Transmission & 1.1 - 1.7 & 4 \\
JWST/NIRCam - Transmission & 2.5 - 4.0 & 5 \\
HST/WFC3 - Emission & 1.1 - 1.7 & 6 \\
JWST/NIRCam F322W2 - Emission & 2.4 - 4.0 & 5 \\ 
JWST/NIRCam F444W  - Emission & 4.0 - 5.0 & 7 \\ 
JWST/MIRI LRS - Emission & 5.0 - 12.0 & 7 \\ \hline
\end{tabular}%
}
\tablefoot{The equilibrium temperature is reported at null Bond albedo. References: [1] \cite{Triaud13}, [2] \cite{Gallet20}, [3] \cite{Mancini14},
[4] \cite{Wong22}, [5] \cite{Bell23}, [6] \cite{Jacobs23}, [7] \cite{Wiser25}.}
\end{table}

\section{Forward model} \label{sec:forward}

\subsection{Interior structure model}
\label{sec:interior_intro}

We adopt the GASTLI open-source package\footnote{\url{https://github.com/lorenaacuna/GASTLI}} \citep[GAS gianT modeL for Interiors;][]{Acuna21,gastli_joss}  as an interior structure model. GASTLI stratifies the planetary interior into two layers: a core (50\% water and 50\% silicate in mass) and an envelope (H/He mixed with water as a proxy for metals). The mass of the core is defined by the mass of the planet, $M_{\rm pl}$ and the core mass fraction (CMF) as $M_{\rm core} = M_{\rm pl} \ \times$ CMF. Together with the planet mass and the CMF, the envelope metal mass fraction, $Z_{\rm env}$, is an input parameter. Thus, the total bulk metal mass fraction of the planet can be computed as $Z_{\rm pl} = \rm CMF + (1-CMF) \times Z_{env}$. GASTLI solves for the pressure $P(r)$, temperature $T(r)$, gravity $g(r)$, and enclosed mass $m(r)$ radial profiles by integrating the differential equations that correspond to hydrostatic equilibrium, an adiabatic temperature gradient, Gauss's theorem and conservation of mass (Eqs. \ref{eq:hydro_eq}-\ref{eqn:mass_conserv})\footnote{$G = 6.6743 \times 10^{-11} \ m^{3} \ kg^{-1} \ s^{-2}$}. The computation of the adiabatic gradient requires the Grüneisen parameter $\gamma$ and the seismic parameter $\phi$, which are calculated using the density $\rho$ and internal energy $E$ (Eq. \ref{eqn:gruneisen}). These two properties are directly obtained from the Equation of State (EOS). The density, internal energy and entropy of H/He are calculated using additive laws, in addition to a correction for non-ideal effects between H and He in the mixture \citep{CD21,HG23}. The properties of the water-rock mixture are derived in a similar manner, by applying the additive laws to the EOS of water \citep{Mazevet2019} and rock \citep{sesame,Miguel22}. The ices accreted by gas giants may contain not only water, but also methane and ammonia. Accounting for non-ideal effects caused by the mixing of rock and ices is challenging because the ice-to-rock mass ratio and the mass fractions of H$_{2}$O, CH$_{4}$ and NH$_{3}$ in the ices are unconstrained \citep{Nettelmann16,MV23}. In the absence of more detailed observational constraints, the use of ideal additive laws remains a reasonable assumption.

%The additive laws are used to calculate the density, internal energy and entropy of the water-rock and H/He-water mixtures. We adopt up-to-date EOSs for H/He , water  and rock  that incorporate non-ideal effects.

\begin{equation} \label{eq:hydro_eq}
    \frac{dP}{dr} = - \rho g
\end{equation}

\begin{equation} \label{eq:gauss_eq}
    \frac{dg}{dr} = - 4 \pi G \rho - \frac{2 G m}{r^{3}} 
\end{equation}

\begin{equation} \label{eq:adiabat_eq}
    \frac{dT}{dr} = - g \frac{\gamma T}{\phi}
\end{equation}

\begin{equation}
\label{eqn:mass_conserv}
\dfrac{dm}{dr} = 4 \pi r^{2} \rho
\end{equation}

\begin{equation}
\label{eqn:gruneisen}
\begin{cases}
\phi = \dfrac{dP}{d \rho}  \\
\gamma = V \  \left(  \dfrac{dP}{dE} \right)_{V} 
\end{cases}
\end{equation}

To solve for the differential equations that govern the internal structure, the following boundary conditions are defined. First, the gravity at the center of the planet is zero, $g(r = 0) = 0$. Secondly, the pressure and the temperature need to be specified at the surface interface. This is the external interface of the outermost computational layer in the 1D grid that represents the radius, not a physical surface as found in rocky planets. These surface conditions, $P(r = R_{\rm int}) = P_{\rm surf}$ and $T(r = R_{\rm int}) = T_{\rm surf}$, are a flexible input for GASTLI, and can be estimated with pre-computed atmospheric grids in an iterative scheme (see Sect. \ref{subsec:gastli_normal_mode}) or can be a free parameter (Sect. \ref{subsec:gastli_joint_mode}). The former is required to obtain a grid of mass-radius-age models for the traditional interior retrievals (Sect. \ref{sec:interior_only_retrievals}), while the latter is more appropriate for our approach to the joint interior-atmosphere retrievals (Sect. \ref{sec:joint_retrievals}). In the joint retrievals, the surface pressure if fixed to $P_{\rm surf} = 1000$ bar.

\subsubsection{Grid of mass-radius-age models}
\label{subsec:gastli_normal_mode}

% Summary of GASTLI's methods (refer the reader to Acuna et al. 2024 and the software paper for more details). This subsection can be split in two parts: (1) the interior and its coupling to a grid of atmospheric models when we have only interior observables (mass, radius, age, [M/H]), which is a recap of the methods in Acuna et al. 2024.

We use GASTLI's default grid of atmospheric models, which was precomputed using petitCODE \citep{Molliere15,Molliere17}. The grid spans a wide range of planetary surface gravities (log $g_{\rm surf}$ = 2.6–4.2 cm/s$^{2}$), equilibrium temperatures (100–1000 K), internal temperatures (50–950 K), metallicities (0.01–250 $\times$ solar), and C/O ratios (0.10–0.55), making it well-suited for gas giants such as WASP-80b. petitCODE computes self-consistent pressure-temperature (P–T) profiles by solving radiative transfer via the correlated-k method, under the assumption of 1D radiative–convective equilibrium. The Bond albedo is calculated self-consistently, based on the energy balance between the emitted and absorbed radiation at the top of the atmosphere. In GASTLI’s default atmospheric grid, chemical equilibrium and cloud-free (clear) atmospheres are assumed. For each coupled model, GASTLI extracts the corresponding P-T profile and computes the atmospheric thickness by integrating the hydrostatic equilibrium equation down to a transit pressure of 20 mbar. This approach enables accurate modeling of observable planetary radii. See \cite{Acuna24_gastli_science_paper} for more details, including the opacities of atmospheric species and their references. 

To achieve consistency between the interior and atmospheric profiles, we use the iterative coupling scheme described in \cite{Acuna21}. 
In this scheme, we define the interior radius ($R_{\rm int}$) as the radius output by the interior model. An initial guess for $R_{\rm int}$ determines the planet's surface gravity, which is then used to interpolate a surface temperature from the atmospheric grid. The interpolation is performed at fixed values of planet mass, envelope metallicity, core mass fraction and internal temperature. The interpolated surface temperature then serves as the boundary condition for the interior model. The new interior radius is recalculated, and the process repeats until convergence in radius and surface temperature is reached. The final mass and radius are calculated consistently by adding its respective atmospheric contributions.

GASTLI calculates the planet’s thermal evolution by solving the entropy ($S$) loss over time ($t$) due to secular cooling. We compute a sequence of models at different internal temperatures ($T_{\rm int}$) -- or thermal states. The luminosity $L$ at each state is related to the internal temperature and total radius, $R_{\rm pl}$, according to Eq. \ref{eqn:lumi}. The luminosity is calculated by also using the Stefan-Boltzman constant, $\sigma = 5.67 \times 10^{-8} \ \rm W \ m^{-2} \ K^{-4}$. Entropy as a function of time, $S(t)$, is obtained by integrating the change in entropy $\partial S$ over time, as defined in Eq. \ref{eqn:therm_evol}. This calculation requires extracting the temperature and enclosed mass profiles, $T(r)$ and $m(r)$, from the interior model to estimate $T(m)$. The resulting $S(t)$ relation allows GASTLI to predict radius and luminosity as functions of age. We adopt GASTLI's default initial entropy, which corresponds to a hot start entropy value $S_{0} = 12 \ k_{\rm B} \ m_{\rm H}$\footnote{$m_{\rm H} = 1.6735 \times 10^{-27}$ kg, and $k_{\rm B} = 1.3806 \times 10^{-23}$ J/K}. For planets whose age $>$ 100 Myr, the choice of the initial entropy does not impact the radius-age and luminosity-age functions \citep[see figures 5 and 9 in][respectively]{SB12,Acuna24_gastli_science_paper}. Moreover, hot-start initial entropy conditions are consistent with observations of cold gas giants \citep{nowak20,trevascus25}.

%is integrated over time using the change of entropy $\partial S$,

%\footnote{$\sigma = 5.67 \times 10^{-8} \ W \ m^{-2} \ K^{-4}$}). 

\begin{equation} \label{eqn:lumi}
    L = 4 \pi \sigma R_{\rm pl}^{2} T_{\rm int}^4
\end{equation}

%\begin{equation} \label{eqn:therm_evol}
%    \frac{\partial L}{\partial m} = - T \frac{\partial S}{\partial t}
%\end{equation}

\begin{equation} \label{eqn:therm_evol}
    \frac{\partial S}{\partial t} = - \frac{L}{M_{\rm pl}} \frac{1}{\int_{0}^{1}  dm \ T(m)}
\end{equation}

% Continue explaining the range of values of this grid for atm. metallicity, mass, internal temperature and core mass fraction. And the fixed value of the equilibrium temperature.

In the mass-radius-age grid, we assume a constant equilibrium temperature at null Bond albedo, equal to that of WASP-80 b, $T_{\rm eq}$ = 825 K. The definition of this parameter is given in Eq. \ref{eq:eq_temp}, where $T_{\star}$, $R_{\star}$ and $a_{\rm d}$ are the stellar effective temperature, stellar radius and orbital semi-major axis (Table \ref{tab:wasp80b_all_obs_data_summary}). petitCODE uses this equilibrium temperature as input and calculates the Bond albedo self-consistently by generating synthetic reflection spectra and integrating over wavelength. We generate a regular grid of GASTLI models 
across a four-dimensional hyperparameter space by varying the planet mass, CMF, the envelope's log(M/H), and internal temperature ($T_{\rm int}$). The input arrays are finely sampled: mass from 0.35 to 0.75 $M_{\rm Jup}$, with $\Delta$M = 0.05 $M_{\rm Jup}$; CMF from 0.0 to 0.90 with $\Delta$CMF = 0.10, plus the value CMF = 0.99 as a last point in the array; log(M/H) is evaluated at -2.0 and 2.4, in addition to the values comprised between 0.0 and 2.0 with $\Delta$log(M/H) = 0.5; $T_{\rm int}$ is calculated at 50 K, and 100 to 800 K with $\Delta T_{\rm int}$ = 100 K. Thus, our grid contains the planetary radius and age for 5670 models in total. This grid will be interpolated with scipy's function \verb|RegularGridInterpolator| in the traditional interior retrievals in Sect. \ref{sec:interior_only_retrievals} to evaluate the forward model.

\begin{equation} 
\label{eq:eq_temp}
    T_{\rm eq} (A_{\rm B} = 0) = \left( \frac{T_{\star}^{4}}{4} \  \left( \frac{R_{\star}}{a_{\rm d}} \right)^{2} \right)^{0.25}
\end{equation}

% If there's room, you can add one figure with a mass-radius plot of the forward models and the mass and radius of WASP-80 b. Say that the expected bulk metal mass fraction in the retrievals is expected to be low from this plot alone.

\subsubsection{Grid for joint retrievals} 
\label{subsec:gastli_joint_mode}

% (2) How we use the interior model in the joint retrievals

% Motivation
As discussed in Sect. \ref{sec:interior_intro}, the traditional interior structure retrievals require the interior model to be coupled to a grid of self-consistent atmospheric models to compute the radius and age as a function of internal temperature. However, for the joint interior-spectrum retrievals, we require the atmospheric profiles to be more flexible than the self-consistent grid. The transmission and emission spectrum may show spectral features that can only be explained by physics that deviate from the assumptions of the self-consistent grid, such as clouds or hazes, or a temperature inversion caused by strong optical absorbers in the upper atmosphere. 

% Coupling
We couple the thermal structure analytical model and the new grid of interior structure models. The coupling algorithm is slightly different from the traditional one described in Sect. \ref{subsec:gastli_normal_mode} and in \cite{Acuna21}. These differences include using additional parameters -- namely the cloud and thermal structure parameters --, an additional step to calculate the mass mixing ratios of the atmospheric species with \textit{easyCHEM}, and computing the transmission and/or emission spectrum with petitRADTRANS after the coupled model has converged (see Sect. \ref{sec:petitradtrans} for a description of petitRADTRANS and \textit{easyCHEM}). For more details on the modified coupling algorithm, see Sect. \ref{sec:coupling_joint}. In the following, we explain how we obtain the grid for the joint retrievals.

%We describe the coupling between the thermal structure analytical model and the new grid of interior structure models in Sect. \ref{sec:coupling_joint}. In the following, we explain how this grid for the joint retrievals is obtained.

% Age
To calculate the grid for joint retrievals, we separate GASTLI's interior module from its atmospheric module. This allows us to calculate the radius ($R_{\rm int}$) and the entropy $S(P=1000 \ \rm bar)$ independently of the atmospheric parameters, such as the atmospheric metallicity log(M/H) or the internal temperature $T_{\rm int}$. 

A set of retrievals in our suite requires to calculate the age because it is taken into account as an observable in the calculation of the log-likelihood (see Sect. \ref{sec:log-likelihood}). Solving for the $T_{\rm int}$-radius-age relations involves computing the function $f_{S} = \partial S/\partial t$ (Eq. \ref{eqn:therm_evol}). In this equation, the luminosity, $L$, is dependent on the intrinsic (or internal) temperature (Eq. \ref{eqn:lumi}). $T_{\rm int}$ is an input to the analytical P-T model that outputs the surface temperature, $T_{\rm surf}$. Consequently, we cannot calculate and store the function $f_{\rm S} = \partial S/\partial t$ in a grid because the internal temperature is not known a priori. Thus, GASTLI's grid for the joint retrievals cannot contain a multidimensional table of $f_{\rm S}$. Instead, we define a new function: $f'_{\rm S} = f_{\rm S} \times L$. Then we store a table for this parameter together with $R_{\rm int}$ and $S(P = $ 1000 bar). For one instance of a converged coupled interior-atmosphere model, we interpolate the internal radius, entropy and $f'_{\rm S}$ values at different surface temperatures that are obtained by varying the internal temperature $T_{\rm i}$ in the \cite{Guillot2010} P-T profile while setting the remaining free parameters constant. We define the internal temperature in Eq. \ref{eqn:lumi} equal to the intrinsic temperature in the P-T profile, $T_{\rm int} = T_{\rm i}$, to compute its corresponding luminosity array. The arrays containing $L(T_{\rm int})$ and $f'_{\rm S}(T_{\rm int})$ are then used to calculate an array for $f_{\rm S}$, being able to solve the differential equation of thermal evolution (Eq. \ref{eqn:therm_evol}). Consequently, we obtain the radius-age curve that allows us to evaluate the age at the $T_{\rm i}$ value proposed by the sampler.

% grid
The input variables for GASTLI's joint grid are the planet mass, CMF, the metal mass fraction of the envelope $Z_{\rm env}$, and the surface temperature and pressure. We set the surface pressure to $P_{\rm surf}$ = 1000 bar, as this is the maximum pressure level appropriate for both our analytical P-T profile and spectrum generator (Sect. \ref{sec:petitradtrans}). Similar to our mass-radius-age grid, we generate a rectangular multidimensional grid by evaluating the interior models at the values specified by fixed arrays of the input parameters $M_{\rm pl}$, CMF, $Z_{\rm env}$ and $T_{\rm surf}$. These arrays consist of $M_{\rm pl}$ = 0.35 to 0.75 $M_{\rm Jup}$, with $\Delta M_{\rm pl}$ = 0.20 $M_{\rm Jup}$; CMF = 0.0 to 0.99, with $\Delta$CMF = 0.10; and $T_{\rm surf}$ = 700 to 6000 K, with $\Delta T_{\rm surf}$ = 100 K. For the envelope mass fraction, $Z_{\rm env}$ = 0.0 to 0.05 at intervals of $\Delta Z_{\rm env}$ = 0.01, continued by the ranges $Z_{\rm env}$ = 0.10 to 0.30 and $Z_{\rm env}$ = 0.40 to 0.80 at intervals of $\Delta Z_{\rm env}$ = 0.05 and 0.10, respectively. Detailed models of the joint grid can be found in Sect. \ref{sec:coupling_joint}.

% Add a figure with one instance of the colormap of the grid, just to show how finely sampled is. You can indicate the exact sampling of the grid in the main text or in a small table.

\subsection{Atmospheric model}
\label{sec:petitradtrans}

% Summary of petitradtrans: radiative transfer, citations for opacities, PT profile, cloud treatment in transmission.

As forward atmospheric model to generate transmission and emission spectra, we use petitRADTRANS \citep[pRT,][]{prt_Molliere2019} version 3.0. \verb|pRT| is an optimized open-source radiative transfer package that allows the user to be flexible in P-T profiles, chemistry, cloud treatment and geometry. The computation of emission spectra requires to obtain the planet-to-star flux ratio, thus we interpolate \verb|pRT|'s in-built library of \textit{PHOENIX} stellar spectra \citep{phoenix_lib_ref} as described by \cite{vanBoekel12}.

% Thermal structure

To parametrize a flexible P-T profile, we employed the simple analytical model provided by \cite{Guillot2010}. It allows us to estimate the thermal structure of irradiated planets by using 4 free parameters in Eq. \ref{eq:PT_guillot10}. The irradiation temperature is defined as $T_{\rm irr} = \sqrt{2} T_{\rm e}$, while the optical depth for a given pressure $P$ is defined as $\tau = P \times \kappa_{\rm IR}/g$. $T_{\rm e}$ is denoted as the equilibrium temperature in the model, $T_{\rm i}$ is the intrinsic temperature, and $g$ is the surface gravity. Finally, $\kappa_{\rm IR}$ corresponds to the mean infrared (grey) opacity, and $\gamma$ is the optical-to-infrared opacity ratio. These parameters can be physically interpreted, for example we could identify $T_{\rm e} = T_{\rm eq} = 825$ K, and $T_{\rm i} = T_{\rm int}$ as calculated by the thermal evolution in the mass-radius-age models (Sect. \ref{subsec:gastli_normal_mode}). However, this would limit the shape of the P-T profile in the atmosphere-only and joint retrievals, so we treat them as free parameters (see Sect. \ref{sec:priors} for their priors). We will only physically interpret $T_{\rm i}$ as the internal temperature $T_{\rm int}$ in the joint retrievals that include the age as an observable to compute the log-likelihood (see Sect. \ref{sec:coupling_joint} and \ref{sec:log-likelihood}).

\begin{multline}
\label{eq:PT_guillot10}
    T^{4} = \dfrac{3 T^{4}_{\rm i}}{4} \left( \dfrac{2}{3} + \tau \right) \\
    + \dfrac{3 T^{4}_{\rm irr}}{4} \left( \dfrac{2}{3} + \dfrac{1}{\gamma \sqrt{3}} + \left( \dfrac{\gamma}{\sqrt{3}} - \dfrac{1}{\gamma \sqrt{3}} \right) e^{-\gamma \tau \sqrt{3} } \right)
\end{multline}

% Chemistry (including absorbers and citations)

We explore both equilibrium chemistry and free chemistry in our suite of retrievals. In equilibrium retrieval, for a given pair of atmospheric metallicity, [M/H], and carbon-to-oxygen ratio, C/O, we calculate the mass fraction abundances by using \textit{pRT}'s pre-computed \textit{easyCHEM} tables \citep{Molliere17,easychem_ref_Paul}. \textit{easyCHEM} calculates and minimises the Gibbs free energy of a mixture as a function of pressure, temperature and elemental makeup following the methods described in the Chemical Equilibrium with Applications (CEA) software \citep{Gordon_cea,Mcbride_cea}. We include a variety of chemical species that are expected in warm gas giant atmospheres to generate spectra in our retrievals. Their respective opacity and line list data references are: H$_{2}$O \citep{H2O_abs_ref}, CO \citep{CO_abs_ref}, CO$_{2}$ \citep{CO2_abs_ref}, CH$_{4}$ \citep{CH4_abs_ref}, NH$_{3}$ \citep{NH3_abs_ref}, HCN \citep{HCN_abs_ref}, Na \citep{Na_abs_ref}, and K \citep{prt_Molliere2019}.

% Clouds

In our retrievals involving transmission spectra, we include condensates as an opacity source. The parametrization of the cloud opacity, $\kappa_{\rm cloud}$, of our condensate model is shown in Eq. \ref{eq:cloud_model}. The cloud model consists of 5 free parameters (see Sect. \ref{sec:priors} for their priors). $\lambda_{0}$ corresponds to the reference wavelength at which the reference opacity, $\kappa_{0}$, is defined. $f_{\rm sed}$ is the cloud scale height decrease factor, and $p$ is the power law coefficient of the opacity with wavelength for $\lambda >> \lambda_{0}$. For pressures higher than the base pressure of the cloud layer ($P_{\rm base}$), the cloud opacity is set to zero as $\kappa_{\rm cloud} (\lambda, P > P_{\rm base}) = 0$. Our condensate model incorporates the behaviour of cloud opacities at different wavelength ranges, which consists of a constant opacity at short wavelengths ($\lambda < \lambda_{0}$) and of a slope at large wavelengths \citep{Dyrek_clouds}.

\begin{equation}
\label{eq:cloud_model}
    \kappa_{\rm cloud} (\lambda,P) = \dfrac{\kappa_{0}}{1+ (\lambda/\lambda_{0})^{p}} \left( \dfrac{P}{P_{\rm base}} \right)^{f_{\rm sed}} \ {\rm if} \ P < P_{\rm base}
\end{equation}

\section{Bayesian framework} \label{sec:bayes}

% Indicate that we use nested sampling - instead of MCMC like previous work (Wilkinson et al.) did. Mention that we use MCMC for the interior observable retrievals only.

In the atmosphere-only and joint retrievals, we adopt nested sampling \citep{multinest_paper1,multinest_paper2,multinest_paper3} to sample the posterior distribution functions as implemented in the \verb|pRT| package by \cite{nasedkin24}. We describe how the log-likelihood function is calculated in these retrievals in Sect. \ref{sec:log-likelihood}. In the traditional interior structure retrievals, we employ the Markov Chain Monte Carlo (MCMC) ensemble sampler implemented in the \textit{emcee} package \citep{emcee}. The likelihood function was constructed following equations 6 and 14 in \cite{Dorn15} and \cite{Acuna21}, respectively. The likelihood function is computed by using the observed values of mass, radius, age, and atmospheric metallicity (see Sect. \ref{sec:interior_only_retrievals}). We initialized 32 walkers and ran the chains for $10^{5}$ iterations. This setup ensures that the chain is sufficiently long to fulfill the MCMC convergence criterion $\tau << N_{\rm chain}/50$ for a typical interior retrieval autocorrelation time of $\tau \simeq$ 100 to 200 \citep{Acuna24_gastli_science_paper}.

\subsection{Likelihood function}
\label{sec:log-likelihood}

% and show Eq. 1 of Nasedkin et al. (prt retrieval JOSS paper). 

%We estimate the likelihood function in our joint interior-atmosphere retrievals using Eq. \ref{eq:likelihood_joint} to estimate the associated $\chi^{2}$ function. 

Eq. \ref{eq:likelihood_joint} shows the $\chi^{2}$ function used to calculate the log-likelihood function (Eq. \ref{eq:norm_likelihood}). The model parameterization vector, $\mathbf{m}$, contains the input variables of a coupled interior-atmosphere model, which are the mass $M_{\rm pl}$ and the other 12 parameters shown in Fig. \ref{fig:joint_coupling_algorithm}. We separate the data in two vectors: the vector containing the spectral fluxes (transmission, emission or both), $ \mathbf{d_{\rm spectra} } $, and the vector containing other observables related to bulk density and evolution information, $ \mathbf{d_{\rm bulk} } $. Depending on the retrieval, $ \mathbf{d_{\rm bulk} } $ may contain the planetary radius (in joint emission retrievals), the planet mass (all joint retrievals) and/or the age. In the retrievals that include the transmission spectra, the planet radius is not included as a bulk observable in $\mathbf{d_{\rm bulk}}$ because this information is already contained in the transmission spectrum. In the first term in Eq. \ref{eq:likelihood_joint}, the vector $s_{i}(\mathbf{m})$ is the spectra binned at the same resolution and central wavelengths as the observed spectra, with corresponding uncertainties $\sigma_{\rm spectra}$. Similarly, in the second term, $g_{j}(\mathbf{m})$ is the coupled interior-atmosphere model parameters related to bulk density and evolution (mass, radius and age), while $\sigma_{\rm bulk}$ are the uncertainties of the observed mass, radius and age. $N_{\rm spectra}$ and $N_{\rm bulk}$ are the number of data points contained in $\mathbf{d_{\rm spectra}}$ and $\mathbf{d_{\rm bulk}}$, respectively.

\begin{multline}
\label{eq:likelihood_joint}
    \chi^{2} (\mathbf{m} \vert \mathbf{d_{\rm spectra}}, \mathbf{d_{\rm bulk}}) = \sum_{i=1}^{N_{\rm spectra}} \dfrac{(s_{i}(\mathbf{m}) - \mathbf{d_{\rm spectra, \ i}})^{2}}{\sigma^{2}_{\rm spectra, \ i}} \\
    + \sum_{j=1}^{N_{\rm bulk}} \dfrac{(g_{j}(\mathbf{m}) - \mathbf{d_{\rm bulk,\ j}})^{2}}{\sigma_{\rm bulk, \ j}^{2}}
\end{multline}

% Previous work by \cite{Wilkinson24} propose to estimate the likelihood function

The final log-likelihood function, $\rm log (\mathcal{L}(\mathbf{m} \vert \mathbf{d_{\rm spectrum}}, \mathbf{d_{\rm bulk}}))$, is obtained by adding a normalization constant to $\chi^{2}$ \citep[Eq. \ref{eq:norm_likelihood}, see also equation 1 in][]{nasedkin24}. This constant takes into account the determinant of the covariance matrix, $\mathbf{C}$, which we assume to be diagonal. In previous work, \cite{Wilkinson24} multiplied the contribution of each data point contained in $\mathbf{d_{\rm bulk}}$ in the second term of Eq. \ref{eq:likelihood_joint} by a weight parameter, $w_{j}$ - in the form $w_{j} \times \dfrac{(g_{j}(\mathbf{m}) - \mathbf{d_{\rm bulk,\ j}})^{2}}{\sigma_{\rm bulk, \ j}^{2}} $. This weight parameter allows to control the importance of an observable in the log-likelihood. In our joint interior-atmosphere retrievals, we choose not to weight the bulk observables (mass, radius and/or age) based on the sensitivity tests we discuss in Sect. \ref{sec:disc_previous}.

\begin{equation}
\label{eq:norm_likelihood}
\rm log (\mathcal{L}) = -\dfrac{1}{2}\chi^{2} - \dfrac{1}{2} log (2 \pi \ \rm det(\mathbf{C}))
\end{equation}

\subsection{Priors}
\label{sec:priors}

\begin{table}[h]
\caption{\label{tab:priors_retrievals} Priors used in the interior and equilibrium chemistry atmosphere retrievals.}
\resizebox{\columnwidth}{!}{%
\begin{tabular}{ll}
\hline
\multicolumn{2}{c}{Interior retrievals}   \\ %\hline
\textbf{Parameter}                     & \textbf{Prior}     \\ \hline \hline
Core mass fraction, CMF $^{a}$      & $\mathcal{U}(0,0.99)$         \\
Atmospheric metallicity, log(M/H) [$\times$ solar]                     & $\mathcal{U}(-2.0,2.4)$         \\
Planet mass, $M_{pl}$ [$M_{\rm Jup}$]                   & $\mathcal{N}(0.55,0.04)$         \\
Age [Gyr]                        & $\mathcal{N}(1.35,0.2)$    \\ \hline
\multicolumn{2}{c}{Atmosphere retrievals} \\ %\hline
\textbf{Parameter}                     & \textbf{Prior}     \\ \hline \hline
log($\kappa_{\rm IR}$) [cm$^{2}$/g] $^{a}$                  & $\mathcal{U}(-6,6)$         \\
$\gamma$  $^{a}$                       & $\mathcal{U}(0.1,2.0)$         \\
$T_{\rm i}$ [K] $^{a}$                            & $\mathcal{U}(0,2000)$         \\
$T_{\rm e}$ [K] $^{a}$                           & $\mathcal{U}(0,2000)$         \\
Planet mass, $M_{pl}$ [$M_{\rm Jup}$] $^{a}$                  & $\mathcal{N}(0.55,0.04)$         \\
Planet radius, $R_{pl}$ [$R_{\rm Jup}$]                   & $\mathcal{N}(0.95,0.03)$         \\
Reference pressure, log$(P_{\rm ref})$ [bar]               & $\mathcal{U}(-6,3)$         \\
Atmospheric metallicity, log(M/H) [$\times$ solar] $^{a}$                    & $\mathcal{U}(-1,3)$       \\
C/O $^{a}$                          & $\mathcal{U}(0.1,1.6)$            \\
log($\kappa_{0}$) [cm$^{2}$/g] $^{a \ b}$               & $\mathcal{U}(-20,20)$            \\
log($P_{\rm base}$) [bar] $^{a \ b}$                  & $\mathcal{U}(-5,3)$            \\
$f_{\rm sed} $ $^{a \ b}$                      & $\mathcal{U}(0,10)$            \\
log($\lambda_{0}$) [$\mu$m] $^{a \ b}$                & $\mathcal{U}(-2,2)$            \\
$p$ $^{a \ b}$                            & $\mathcal{U}(0,6)$  \\
offset$_{\rm dataset}$ [ppm] $^{a}$                            & $\mathcal{N}(0,100)$  \\
Stellar radius, $R_{\star}$ [$R_{\odot}$] $^{a}$                            & 0.571 (fixed)  \\  
Stellar effective temperature, $T_{\star}$ [K] $^{a}$                          & 4145 (fixed)  \\ \hline
\end{tabular}%
}
\tablefoot{$\mathcal{N}(\mu,\sigma)$: Gaussian prior. $\mathcal{U}$(low, high): uniform prior. $^{a}$ The joint interior-atmosphere retrievals use similar priors for this parameter. $^{b}$ Cloud parameters apply only when transmission spectra data are considered in the retrieval.}
\end{table}

% Add also the priors that were considered in the spectrum and the joint retrievals. 

Table \ref{tab:priors_retrievals} indicates the prior distributions used for the free parameters in the interior, and equilibrium chemistry atmosphere retrievals. In the interior retrievals, if the atmospheric metallicity is included as an observable in the log-likelihood, the prior on log(M/H) is not uniform (see R3-R6 in Sect. \ref{sec:interior_only_retrievals}). Instead, we use a normal prior whose mean and standard deviation are informed by the imposed value.

For free chemistry retrievals, we assume the same priors of the equilibrium chemistry retrievals except for log(M/H) and C/O since these two parameters do not apply. The mass fraction abundances of the chemical species are free parameters with a prior log$(X_{\rm species}) = \mathcal{U}(-10,0)$. For consistency, the joint interior-atmosphere retrievals use similar priors for those parameters that are relevant when bulk density data, age and spectra are combined as observables (see caption in Table \ref{tab:priors_retrievals}). The reference pressure does not exist as prior in the joint retrievals because it is fixed to $P_{\rm ref}$ = 1000 bar, where the interior-atmosphere interface is located. Similarly, as the planet radius is computed by the interior grid consistently, it does not require a prior in the joint retrievals. Finally, Table \ref{tab:priors_retrievals} shows that the prior on internal temperature is uniform between 0 and 2000 K. In the joint retrievals that include age as an observable, this prior is changed to $\mathcal{U}$(0,500) K to improve convergence. This prior choice is further motivated by our self-consistent interior-atmosphere models (see Fig. \ref{fig:interior_forward_nocore}), which show that WASP-80 b's internal temperature does not exceed a few hundred Kelvin in the absence of extra heating sources. The age of WASP-80 b is approximately 1 Gyr, which corresponds to internal temperatures $T_{\rm int}$ = 100-300 K according to forward models, even in cloudy atmospheres \citep{Poser19,Poser24}. Internal temperatures above 900 K hinder convergence towards older ages in our joint retrievals, as they restrict the explored age range to 1-10 Myr.

Our retrievals also include offset parameters between a reference data set and the other data sets in the same geometry. Specifically, the NIRCam data set is our reference in transmission, introducing two free offsets between the NIRCam and STIS, and the NIRCam and WFC3 data sets. In emission, our reference data set is NIRCam F322W2. Thus, we use three free offsets: one for WFC3, one for NIRCam F444W, and one for MIRI LRS. These offsets account for differences in instruments and data reductions, for which we adopt Gaussian priors, $\rm offset_{\rm data set} = \mathcal{N}(0,100)$ parts per million (ppm) in transit depth and planet-to-star flux in transmission and emission, respectively. Previous work report offsets of $\sim$200 ppm between JWST and HST data sets \citep{Lothringer25}, so a normal distribution with a standard deviation of 100 ppm is a reasonable offset prior in our retrievals.

\section{Atmosphere spectrum retrievals}
%\section{Transmission spectrum retrievals}

\subsection{Transmission spectrum retrievals}
\label{sec:transmission_only}

% General: First results section. Summary of retrievals that use the transmission spectrum only. These are two retrievals: one with equilibrium chemistry and one with free chemistry. 

% Part 1 in this section: Composition - log(Fe/H) and C/O ratio

% Equilibrium chemistry

%in our equilibrium chemistry retrieval

\begin{figure}[h]
   \centering
   \includegraphics[width=\columnwidth]{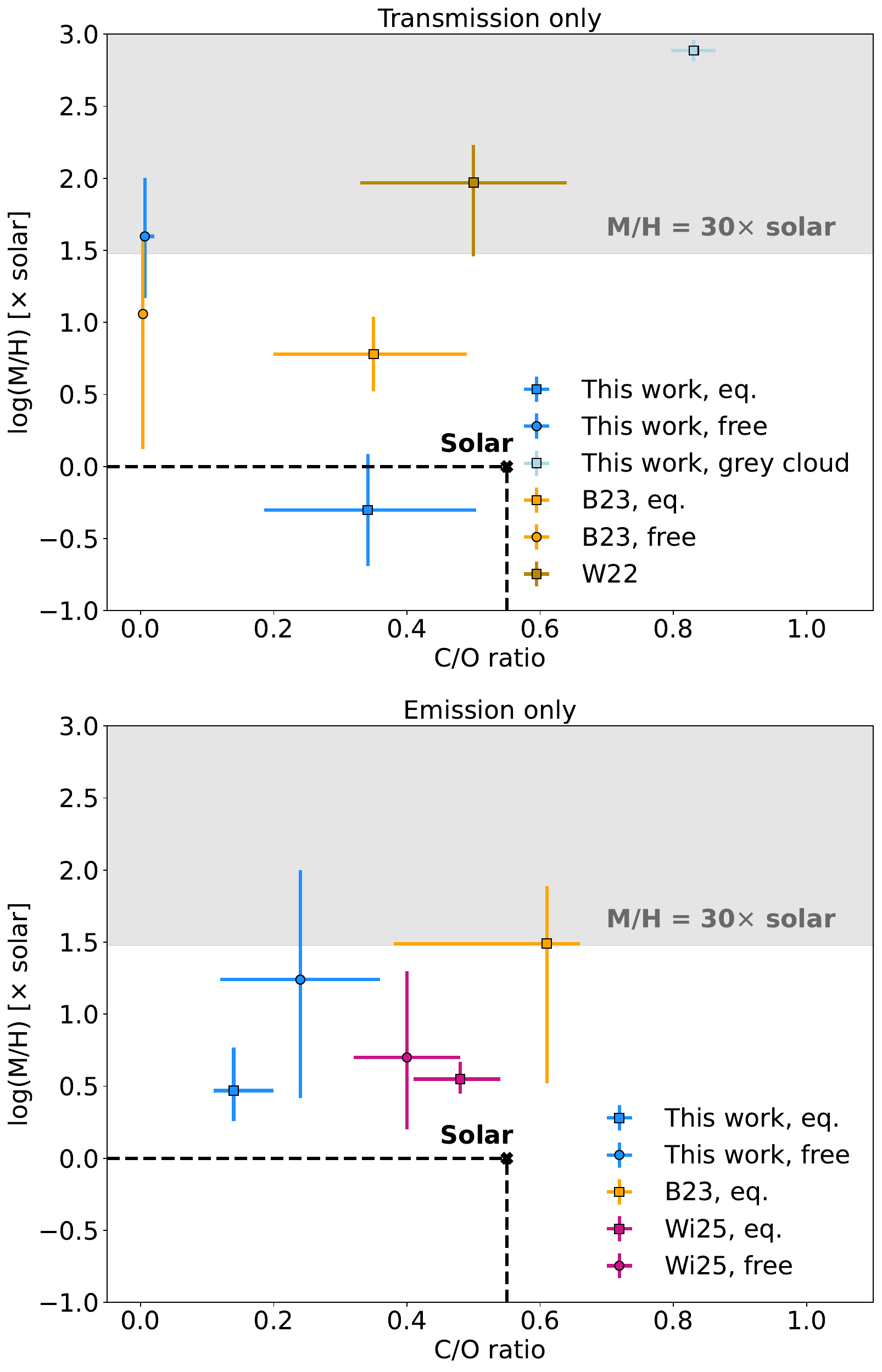}
      \caption{Comparison between our atmospheric composition estimates and previous work from transmission-only (upper panel) and emission-only (lower panel) retrievals. Previous work include \cite{Bell23} (B23, orange), \cite{Wong22} (W22, dark gold) and \cite{Wiser25} (Wi25, violet). The shaded area indicates the region forbidden by our interior-only retrievals at 4$\sigma$ (see Sect. \ref{sec:interior_only_retrievals}).}
      
         \label{fig:transmission_retrieval_summary}
\end{figure}

As a starting point for our transmission-only retrievals, we fit an equilibrium chemistry model. In our retrieval, WASP-80 b's atmospheric composition ranges from sub-solar to solar metallicity, and it is consistent with a solar C/O = 0.55 within 1.3$\sigma$.  This retrieval obtains an atmospheric metallicity of log(M/H)$_{\rm eq}$= $-0.32^{+0.41}_{-0.36}$ and (C/O)$_{\rm eq}$ = 0.33$^{+0.17}_{-0.14}$. Our C/O ratio agrees well with similar equilibrium chemistry retrievals by \cite{Wong22} (W22) and \cite{Bell23} (B23), whose 1$\sigma$ estimates span C/O = 0.20 - 0.60 (see Fig. \ref{fig:transmission_retrieval_summary}). In contrast, the equilibrium retrievals by previous work suggest that the atmospheric metallicity is super-solar, being M/H $\simeq$ 3-10 $\times$ solar (B23) and M/H $\simeq$ 25-100 $\times$ solar (W22). In summary, our equilibrium chemistry retrieval agrees with previous work on C/O ratio, but obtains a solar-like metallicity in contrast to their super-solar estimates.

% Free chemistry 
We also explore free chemistry retrievals to constrain the abundances of individual absorbing species. We obtain tight constraints on the mass fraction abundances of H$_{2}$O and CH$_{4}$, log($X_{\rm H_{2}O}$)$_{\rm free}$ = $-0.61^{+0.38}_{-0.37}$ and log($X_{\rm CH_{4}}$)$_{\rm free}$ = $-2.99^{+0.53}_{-0.56}$, respectively. CO and CO$_{2}$ abundances are poorly constrained, spanning more than four orders of magnitude both in our retrievals and B23. Our estimate for CH$_{4}$ agrees extremely well with that obtained by B23 in their free chemistry retrievals. Our retrieved H$_{2}$O mass fraction ratio is consistent within 1$\sigma$ with B23's water abundance estimate.

%We perform a free chemistry retrieval considering only the NIRCam dataset to test whether this difference in water abundance uncertainties could be caused by their lack of data at $\lambda = $ 0.4-1.7 $\mu$m. \textcolor{red}{In this retrieval, our water abundance uncertainties are reduced by a factor of two compared to our nominal free chemistry retrieval.} This demonstrates that the observational data have an effect on the free water abundance uncertainties, but it is unlikely to be the only source of the difference. Our 

Our free retrieval contribution function indicates that the abundances are probed at a pressure $P_{\rm transit} \simeq 10^{-3}$ bar. Thus, we compare our free retrieved abundances to the expected abundances at the transit pressure from our equilibrium chemistry retrieval. We find that the free CH$_{4}$ abundance is compatible with its expected equilibrium abundance within $<$2$\sigma$. In contrast, water is 4$\sigma$ more abundant in the free retrieval than in equilibrium.

%\textcolor{red}{Nonetheless, this discrepancy is reduced to 2$\sigma$ if we compare to the free chemistry water abundance reported by B23, given their larger uncertainties.} 

We estimate the overall atmospheric metallicity and C/O ratio from our free retrieved abundances, and obtain log(M/H)$_{\rm free}$ = $1.60^{+0.41}_{-0.43}$ and (C/O)$_{\rm free}$ = 0.026$^{+0.005}_{-0.023}$ (shown in Fig. \ref{fig:transmission_retrieval_summary}). This estimate agrees with the scenario presented by B23 in their free retrievals, in which WASP-80 b's atmosphere is super-solar and has a very low C-to-O ratio with log(M/H)$_{\rm B23, \ free}$ = 1.06$^{+0.55}_{-0.94}$ and (C/O)$_{\rm B23, \ free}$ = 0.004$\pm$0.002.

% Figures: (1) plot comparing all the C/Os and [M/H]s; (

% Discuss the tension between these two and previous work (Bell+, Wong+). Our equilibrium retrieval is subsolar, but previous work and our free retrieval indicate [M/H] goes from 10 to 100 x solar. 

% Part 3: Bayesian evidence and clouds discussion included here too

% \usepackage{graphicx}
\begin{table}[h]
\caption{\label{tab:fits_transmission_only} Goodness-of-fit metrics for the three transmission spectrum retrievals.}
\resizebox{\columnwidth}{!}{%
\begin{tabular}{lccc}
\hline
Transmission retrieval model                         &  $k$ & $\chi^{2}_{\rm reduced}$ & ln($Z$) \\ \hline \hline
(1) Equilibrium chemistry, $\kappa_{\rm cloud}$ model & 16 &  0.75   &  1095.7     \\
(2) Free chemistry, $\kappa_{\rm cloud}$ model  & 24  & 0.78   &  1089.8    \\
(3) Equilibrium chemistry, grey cloud & 13 &  0.94   &    1082.9  \\ \hline
\end{tabular}%
}
\tablefoot{The reduced $\chi^{2}$ is calculated using the degrees of freedom, DoF = $N_{\rm spectra}-k$, where $k$ is the number of free parameters in the retrieval. ln($Z$) corresponds to the Bayesian evidence.}
\end{table}

We adopt the equilibrium chemistry retrieval as our fiducial transmission-only model as it has the highest Bayesian evidence (see Table \ref{tab:fits_transmission_only}). Our free chemistry retrieval and previous analyses indicate a super-solar metallicity, whereas our fiducial model suggests a sub-solar to solar composition. We explore whether this discrepancy could be explained by different cloud treatments by performing a new retrieval. This retrieval has in common with our fiducial model the equilibrium chemistry, but uses a cloud model that is 
different from our $\kappa_{\rm cloud}$ model (Eq. \ref{eq:cloud_model}). This is a simple cloud model that consists of a gray cloud deck with two free parameters. The first parameter corresponds to the pressure at which the cloud deck is located, $P_{\rm cloud}$. This means that the cloud opacity is completely opaque at all wavelengths for $P > P_{\rm cloud}$, being similar to the cloud model assumed in B23. The second free parameter constitutes a haze factor, which scales the Rayleigh scattering opacity characteristic of hazes. With this cloud parameterization, we obtain an atmospheric metallicity and a C-to-O ratio that are super-solar: log(M/H)$_{\rm grey \ cloud}$ = 2.89$^{+0.06}_{-0.08}$ and (C/O)$_{\rm grey \ cloud}$ = 0.84$^{+0.03}_{-0.04}$. Thus, we demonstrate that for the particular case of WASP-80 b, the cloud treatment has a significant effect on the estimated composition, particularly the atmospheric metallicity and C/O ratio. Previous work also found that cloud parametrization can impact the inference of atmospheric composition \citep{Line16,Macdonald17,Fisher18,Lueber23,RoyPerez25}.

%We note that the analyses by previous work -- both in equilibrium and free chemistry -- and our free chemistry retrieval indicate a super-solar metallicity, whereas our equilibrium chemistry retrieval suggests a sub-solar to solar composition. We explore whether this discrepancy could be explained by different cloud treatments between our retrievals and previous work by carrying our nominal equilibrium retrieval with a different cloud model.  

We compare the cloud properties between our nominal equilibrium and free retrievals. All cloud parameters are similar except for the cloud base pressure, being the cloud parameter with the narrowest constraints. The equilibrium retrieval tends to locate the cloud layer at pressures higher than the transit pressure, $P_{\rm base, \ eq}$ = 0.1 to 100 bar (1$\sigma$) whereas the free chemistry retrieval locates the cloud base pressure at higher altitudes that are consistent with the transit pressure, $P_{\rm base, \ free} = 1.5 \times 10^{-3}$ to 3 bar. This suggests that clouds are a greater source of absorption in the free retrieval than the equilibrium one. Thus, in equilibrium chemistry, the opacity is dominated by the species absorption, namely CH$_{4}$, as the C/O ratio is greater than the free retrieval estimate by $\sim$0.4. 

Finally, we compare the Bayesian evidence between the three models, (1) equilibrium chemistry, (2) free chemistry, and (3) equilibrium chemistry, with a grey cloud deck. The equilibrium chemistry model is preferred over the free chemistry one by $ln(B_{12}) = ln(Z_{1}) - ln(Z_{2})$ = 5.9 and the grey cloud deck equilibrium model by $ln(B_{13})$ = 12.8. The comparison of these models is not as straightforward as comparing nested testing to estimate the detection confidence of a particular molecule. In this case, the degrees of freedom is different between models, where the most complex models have more free parameters (see Table \ref{tab:fits_transmission_only}), and the comparison can be more sensitive to the choice of priors. Under this word of caution, we compare the Bayes factors of the two comparisons and estimate lower and upper limits for their confidence levels according to \cite{Schmidt25} and \cite{Trotta08}, respectively \citep[see also][for a detailed discussion on how to avoid overestimation of confidence levels in Bayesian model comparisons]{Kipping_Benneke}. The first comparison, $ln(B_{12})$ = 5.9 is equivalent to a confidence level of 1.5$-$4$\sigma$. Similarly, the second comparison $ln(B_{13})$ = 12.8 corresponds to 1.5$-$5$\sigma$. Even if it is at low confidence, a sub-solar [M/H] with a solar C/O ratio and wavelength-dependent cloud extinction at low altitudes is preferred over a scenario with super-solar [M/H] with very low C/O ratios and high altitude clouds (see Fig. \ref{fig:transmission_retrieval_bestfit} for the best fit and residuals).

%\section{Emission spectrum retrievals}
\subsection{Emission spectrum retrievals}
\label{sec:emission_only}

% Second results section. Summary of retrievals that use the emission spectrum only. These are two retrievals: one with equilibrium chemistry and one with free chemistry. Both agree well in the PT profile (expected) - in composition, we only obtain constraints on the CH4 abundance. 

% Compare to Bell+ emission retrieval. 

% Chemistry

In the following, we explore retrievals for the emission spectrum only. Our emission equilibrium chemistry retrieval estimates an atmospheric metallicity of log(M/H)$_{\rm emission, \ eq}$ = $0.47^{+0.30}_{-0.21}$ and (C/O)$_{\rm emission, \ eq}$ = 0.14$^{+0.06}_{-0.03}$. The emission-only free chemistry retrieval agrees well within 1$\sigma$ with the emission-only equilibrium estimate, log(M/H)$_{\rm emission, \ free}$ = 1.24$_{-0.82}^{+0.76}$ and (C/O)$_{\rm emission, \ free}$ = 0.24$\pm 0.12$. Overall, the emission-only retrievals agree well in C/O ratio with our transmission-only retrievals. In contrast, the atmospheric metallicity in emission-only retrievals tends to be slightly super-solar, agreeing well with the free chemistry transmission retrieval (see both panels in Fig. \ref{fig:transmission_retrieval_summary}). 

We estimate the Bayes factor between the equilibrium and free chemistry emission-only retrievals: $ln(B) = ln(Z_{\rm free})-ln(Z_{\rm equilibrium}) = 1538.1 - 1537.0 = 1.1$. This value of Bayes factor is well below 1$\sigma$, meaning that the emission spectrum alone does not show a preference for equilibrium over free chemistry and viceversa.

Fig. \ref{fig:transmission_retrieval_summary} (lower panel) shows a comparison of our emission-only estimates and those of previous work. All five estimates agree well within uncertainties in an atmospheric metallicity that ranges between solar and 100$\times$ solar. The equilibrium retrievals in \cite{Wiser25} and our work agree that the atmospheric metallicity is [M/H]$\sim$3-10$\times$ solar. However, our retrievals constrain C/O $<$ 0.40, while \cite{Bell23} and \cite{Wiser25} obtain 1$\sigma$ intervals C/O = 0.40-0.65. Our lower C/O estimate could be due to our assumption of equilibrium chemistry, while \cite{Wiser25} consider a free vertical eddy diffusion coefficient, $K_{zz}$ in their retrievals. Thus, their grid may allow for atmospheric models that have low abundances of carbon-bearing molecules (CH$_{4}$, CO$_{2}$, CO) at moderately high $K_{zz}$ values without requiring low C/O ratios. In addition, we report a weak correlation between C/O ratio and the offset between the MIRI LRS and NIRCam F322 data sets. For a zero value of this offset, we observe a higher C/O$\sim$0.4. \cite{Wiser25} do not use offsets between the NIRCam and MIRI data sets, which may bias the C/O value to slightly higher estimates.

%While assuming equilibrium,  

%This C-to-O ratio is compatible with a solar value, and agrees with B23 on C/O $<$ 0.65. Of all atmospheric species, the emission free retrieval can only obtain tight constraints on the mass fraction abundance of methane with log($X_{CH_{4}}$)$_{\rm emission, \ free}$ = -3.80$^{+0.99}_{-0.60}$, which is in good agreement within uncertainties with our free transmission retrieval (Sect. \ref{sec:transmission_only}). 

% PT profile

\begin{figure}[h]
   \centering
   \includegraphics[width=\columnwidth]{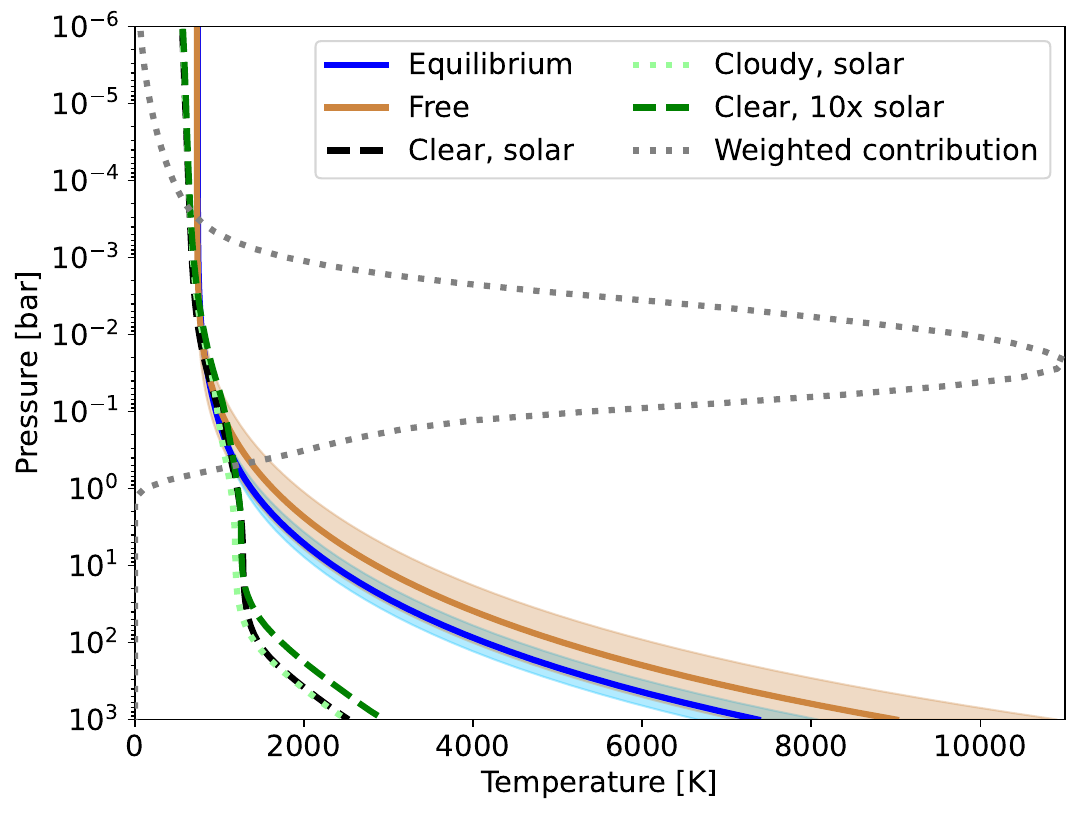}
      \caption{Pressure-Temperature (PT) profile of WASP-80 b as constrained by our two emission-only retrievals. The shaded region indicates the 1$\sigma$ area of the retrieval, while solid lines correspond to the mean. We show self-consistent 1D models from \cite{Molliere17} for comparison, including cloudy model 1 in their table 2 (see text). The grey dotted line shows the contribution function from our emission-only retrievals.}
         \label{fig:PT_profile_emission}
\end{figure}

The emission retrievals also allow us to determine the thermal structure at high altitudes. Fig. \ref{fig:PT_profile_emission} shows the 1$\sigma$ confidence regions estimated by both emission-only retrievals. The spectrally weighted contribution function is higher at pressures where the emission spectrum is more sensitive to the thermal structure, indicating the pressures that are probed by the data. These pressures are $P = 10^{-3}-1$ bar, for which the thermal structure is well constrained with $T$ = 700$-$1000 K. These temperature range is in excellent agreement with that estimated by \cite{Bell23} and \cite{Wiser25}. 

We overplot self-consistent models in our P-T diagram to compare to the retrieved profiles. These models correspond to the clear models provided by \cite{Molliere17}, which agree well with the retrieved profiles at the pressures probed by the emission data. In addition, we show one cloudy model that agrees well, which is model 1 in table 2 from \cite{Molliere17}. This cloudy model assumes irregular grains via the distribution of hollow spheres (DHS), a standard settling parameter $f_{\rm sed} = 3$, and includes iron (Fe) clouds. We rule out cloud models with $f_{\rm sed}<1$ and atmospheric cloud mass fractions $X_{\rm max} > 3 \times 10^{-5}$, as their temperatures (not shown) at pressures $P=1-0.1$ bar and $P=10^{-2}-10^{-3}$ bar are significantly colder and warmer, respectively, than our retrieved profiles. 

%agree well with the self-consistent clear models at solar and 10$\times$ solar compositions. 

%The self-consistent cloudy model is a few hundred K colder than the retrieved temperature at $P=$ 0.1-1 bar,

%which may be indicative of a radiative layer at those pressures in WASP-80 b.

%The temperature in the upper atmosphere is isothermal and slightly warmer than the clear, solar model. Clouds and the dayside average bring the upper temperature closer to the retrieval estimate.

%For atmospheres in chemical equilibrium, CH$_{4}$ is abundant, while at high deep temperatures it is depleted \citep{Moses13,Fortney20,Sing_24_WASP107b}. Thus, a low temperature at the bottom of WASP-80 b in chemical equilibrium retrievals is likely caused by the need to produce CH$_{4}$ as an absorber at higher altitudes, as the emission spectrum is sensitive to its abundance. 

%For pressures $P >$ 1 bar, the temperature is very degenerate, ranging between $T(P>1 \ \rm bar)$ = 1000-6000 K. This is expected as the pressures probed in emission are low. 

% Figures: (1) plot comparing the PT profiles; 

% This plot is going to be merged in Fig. 4 with the transmission spectrum. (2) plot of the nominal fit + spectrum. Mention chi square to prove it is a good fit.

\section{Interior retrievals}
\label{sec:interior_only_retrievals}

\begin{table*}[h]
\caption{\label{tab:interior_only_summary} Summary of mean and uncertainties of retrieved compositional parameters in the interior-only retrievals.}
%\resizebox{\columnwidth}{!}{%
\centering
\begin{tabular}{lccccc}
\hline
Observables, $\mathbf{d_{\rm bulk}}$ & CMF & log(M/H) & $Z_{\rm env}$ & $Z_{\rm planet}$ & $ \vert \rm log(M/H)_{\rm mean \ PDF} - \rm log(M/H)_{\rm obs} \vert$ \\ \hline \hline
R1: $\{M_{\rm pl}, R_{\rm pl} \}$ &  0.08$^{+0.07}_{-0.05}$   &  $-0.61^{+0.91}_{-0.94}  $  &   0.01$^{+0.03}_{-0.01}$     &   0.10$^{+0.07}_{-0.05}$  & - \\
R2: $\{M_{\rm pl}, R_{\rm pl}$, Age$ \}$ &  0.12$\pm$0.06   &   $-0.47^{+1.01}_{-1.04}$      &   0.01$^{+0.06}_{-0.01}$     &   0.15$\pm$0.05  & -  \\
R3: $\{M_{\rm pl}, R_{\rm pl}$, log(M/H)$^{a}\}$ &    0.09$^{+0.07}_{-0.05}$ &    $-0.31^{+0.37}_{-0.36}$      &   0.01$^{+0.01}_{-0.00}$     &  0.10$^{+0.07}_{-0.05}$  &  0.025 $\sigma$  \\
R4: $\{M_{\rm pl}, R_{\rm pl}$, log(M/H)$^{b}\}$ &    0.06$^{+0.16}_{-0.05}$ &   0.92$_{-0.27}^{+0.30}$       &   0.11$^{+0.11}_{-0.03}$     &  0.17$^{+0.25}_{-0.05}$  &   1.70 $\sigma$ \\
R5: $\{M_{\rm pl}, R_{\rm pl}$, Age, log(M/H)$^{a}\}$ &  0.14$\pm$0.05   &   $-0.29^{+0.39}_{-0.37}$       &   0.01$^{+0.01}_{-0.00}$     &  0.16$\pm$0.05   &  0.05 $\sigma$   \\
R6: $\{M_{\rm pl}, R_{\rm pl}$, Age, log(M/H)$^{b}\}$  &  0.06$^{+0.06}_{-0.04}$   &  0.90$^{+0.14}_{-0.23}$        &    0.11$^{+0.03}_{-0.02}$    &  0.16$\pm$0.04  &    1.72 $\sigma$  \\ \hline
\end{tabular}%
%}
\tablefoot{CMF, $Z_{\rm env}$ and $Z_{\rm planet}$ are non-observable parameters, while log(M/H) is an observable parameter we fit for in R3-R6. The metric $ \vert \rm log(M/H)_{\rm mean \ PDF} - \rm log(M/H)_{\rm obs} \vert $ quantifies how the mean of the posterior distribution function of the retrieved log(M/H) differs from the mean observational value. $^{a}$ We adopt the atmospheric metallicity value from our transmission-only equilibrium chemistry retrieval (sub-solar), log(M/H) = $-0.32^{+0.41}_{-0.36}$. $^{b}$ We assume an atmospheric metallicity value from our transmission-only free chemistry retrieval (super-solar), log(M/H) = 1.60$^{+0.41}_{-0.43}$.}
\end{table*}

% Third results section. Summary of retrievals with interior observables only. These are 6 in total. Talk about: 

We run a suite of six interior-only retrievals that take into account the observables traditionally used to characterize planetary interior structure. To assess the effect of each observable, we vary the number of observables included in the retrieval and, in the case of the atmospheric metallicity, explore the effect of its assumed value. The atmospheric metallicity is not incorporated into the interior retrieval by including the spectrum in the log-likelihood -- this approach is applied in Sect. \ref{sec:joint_retrievals} via Eq. \ref{sec:log-likelihood}. Instead, we treat atmospheric metallicity as an independent observable parameter and fit it alongside mass, radius and age using a log-likelihood function with a prior. We consider two scenarios for the atmospheric metallicity: a sub-solar value derived from our transmission-only, equilibrium chemistry retrieval (fiducial), and a super-solar value from our transmission-only, free chemistry retrieval. Table \ref{tab:interior_only_summary} summarizes the observables and shows the mean and uncertainties of the retrieved compositional parameters. 

%The last column displays the difference in $\sigma$ between the mean of the posterior distribution function (PDF) and the observed mean value.

% MOVE THIS TO THE APPENDIX? Add 1 summary table (ok, this is already done) with mean and uncertainties of posterior + (if room) 1 figure with corner plot of nominal retrieval (mass, radius, age, subsolar [M/H]). If not enough room, this goes in the appendix.}

% (2) Tension between free transmission retrieval and interior retrieval with mass, radius, age and [M/H] - this tension is resolved if [M/H] < 1 (subsolar). 

% Close this paragraph with a conclusion: WASP-80 b cannot have an atmospheric metallicity greater than 10 x solar, unless an unknown mechanism is inflating the planet to a very high Tint (how much?), which is equivalent to a temperature at x bar of x K. Cliffhanger: discuss inflation mechanisms for WASP-80 b in Sect. 9 (in the dilute core part)

In the retrievals that use a super-solar atmospheric metallicity (R4 and R6), the posterior distribution of the atmospheric metallicity does not agree with the imposed prior distribution. To quantify this discrepancy, we define the metric $ \vert \rm log(M/H)_{\rm mean \ PDF} - \rm log(M/H)_{\rm obs} \vert $, which quantifies the difference between the posterior mean and the observational mean, expressed in prior standard deviation units (or $\sigma)$. A value of $ \vert \rm log(M/H)_{\rm mean \ PDF} - \rm log(M/H)_{\rm obs} \vert > 1 \sigma$ indicates that R4 and R6 have difficulty fitting log(M/H) and the other observables simultaneously. This tension arises because the density of WASP-80 b is lower than that of a core-less (CMF = 0) planet with a super-solar envelope composition.

In the following, we discuss the maximum atmospheric metallicity allowed by the density of WASP-80 b. R6 shows that if the age is taken into account, the maximum possible metallicity is $\sim$10 $\times$ solar. This is consistent with the mass-radius-age forward models where a planet with WASP-80 b's mean mass and log(M/H) = 1 requires to have no core (CMF = 0) to match its observed radius at its current age (see Fig. \ref{fig:interior_forward_nocore}). In the absence of external heat sources and a non-homogeneous interior in WASP-80 b, an atmospheric metallicity that is consistent with the bulk density cannot be [M/H]$>$10$\times$ solar at 1$\sigma$, and [M/H]$>$30$\times$ solar at 4$\sigma$. Thus, there is a discrepancy in [M/H] between our interior-only analysis, and our transmission-only, free chemistry retrieval as well as the estimates presented in previous work. We further discuss how free joint interior-atmosphere retrievals show degeneracies between envelope composition and chemistry in Sect. \ref{sec:joint_retrievals}.

% THIS IS A DIFFERENT PARAGRAPH - I move this to the discussion (heating mechanisms)
% If the envelope metallicity is increased to 30 $\times$ solar in a core-less scenario, we require a net internal temperature $T_{\rm int}$ = 600 K to match WASP-80 b's current density. Given its age, its current internal temperature -- due to homogeneous, secular cooling with no extra heat sources -- is $T_{\rm int} \simeq$ 100 K, meaning that its luminosity would have to be increased by a factor of $\sim$1300 to match the observational data if [M/H] $\geq 30 \times$ solar. According to our grid of self-consistent \textit{petitCODE} atmospheric models, this is equivalent to an interior-atmosphere interface temperature of $T (10 \rm \ bar) = 2300$ K, which corresponds to $T_{\rm surf} (1000 \ \rm bar) \sim$ 4000 K if a dry, convective adiabat is extended to 1000 bar. This high-temperature scenario could be possible if WASP-80 b had external heat sources or a non-homogeneous interior. We discuss possible inflation mechanisms for WASP-80 b and the likelihood of a hot, non-homogeneous interior in Sect. \ref{sec:hightemp_disc}. 

The retrievals that do not incorporate the atmospheric metallicity as an observable (R1 and R2), and those that use its sub-solar value (R3 and R5), can reproduce all observables within the 1$\sigma$ uncertainties. This means that the retrieval can find forward models that are physical and fit the values imposed by the observational data. These four retrievals agree well in their retrieved compositional parameters, suggesting that the core mass fraction of WASP-80 b is CMF $\sim$ 0.03$-$0.19 at 1$\sigma$. This CMF range is equivalent to a core mass $M_{\rm core}$ = 5.2$-$33.2 $M_{\oplus}$. Its envelope metal mass fraction is low, with $Z_{\rm env}< 0.07$ within 1$\sigma$. Thus, the total bulk metal mass fraction of WASP-80 b ranges between 0.10 and 0.15 with standard deviations of 0.05. We compare this estimate to the joint retrieval in Sect. \ref{sec:joint_retrievals} and discuss its implications for WASP-80 b's formation in Sect. \ref{sec:formation_disc}.

%$\sim$5 weight percentage (wt\%)

% This can already be expected from the most degenerate retrieval in the suite, R1, where the maximum atmospheric metallicity is log(M/H) = 0.30 at 1$\sigma$ ($\sim$2 at 3$\sigma$).

% ----------------

% ELABORATE MORE ON THIS IN THE DISCUSSION: What is the final bulk metal content of WASP-80 b? This is answered briefly in the last paragraph of this section - however, I'd like to discuss what this means for planet formation in the discussion section.

% MOVE THIS TO THE DISCUSSION? (1) uncertainties for mass, radius and age, and inference of CMF and Zplanet. Would improving the uncertainties of these improve the composition constraints? 

% (3) Confirm what you discussed in Acuna et al. 2024: having the atmospheric metallicity is not constraining if the age's uncertainties are large. 

\section{Joint interior-atmosphere retrievals}
\label{sec:joint_retrievals}

\begin{figure*}[h]
   \centering
   \includegraphics[width=\textwidth]{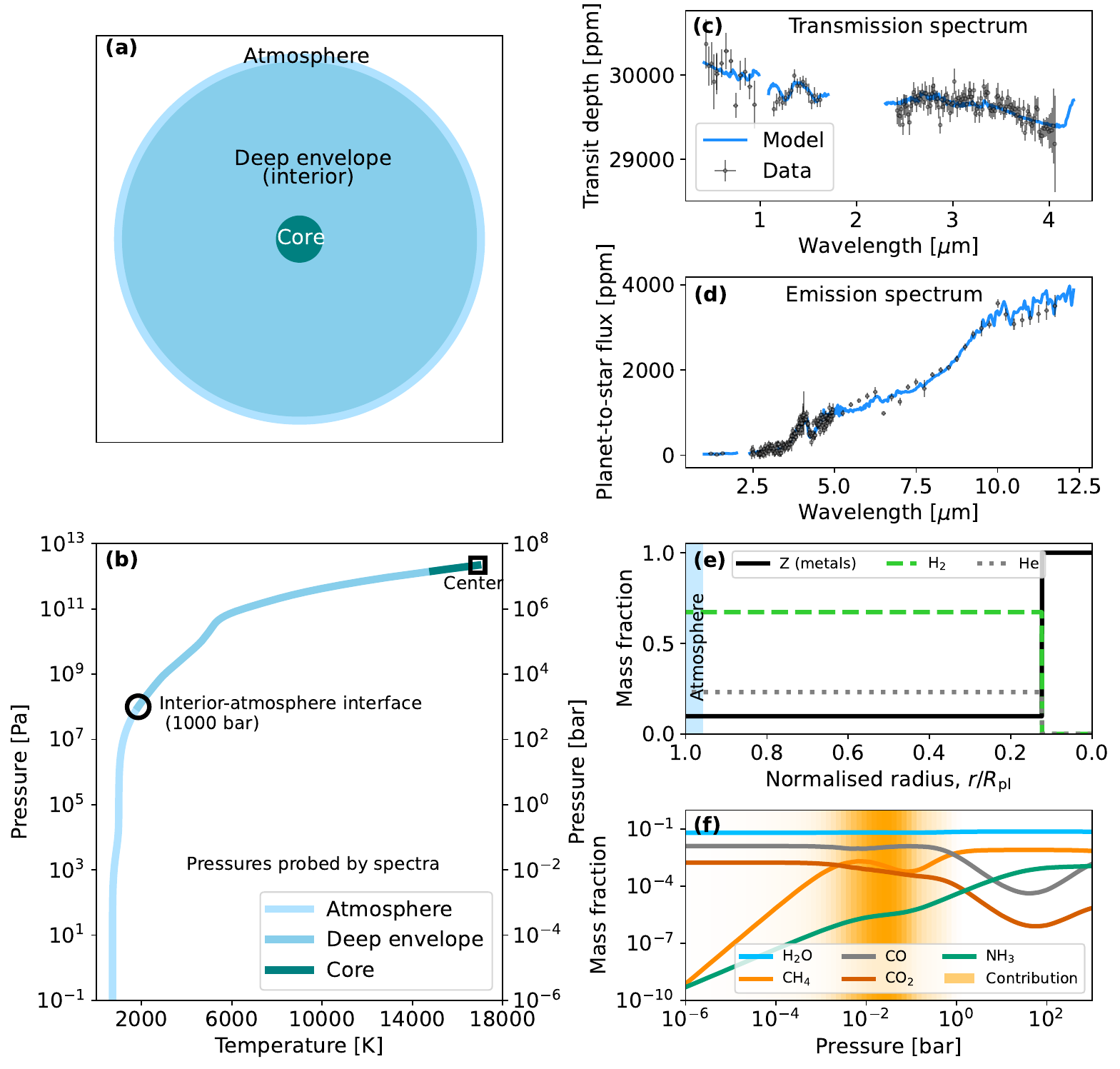}
      \caption{Summary of the mean forward model from one of our two fiducial interior-atmosphere joint retrievals, JR6, of WASP-80 b. This model corresponds to a core mass fraction (CMF) of 0.02, an atmospheric metallicity of M/H = 2.75 $\times$ solar, C/O = 0.12, and $T_{\rm int}$ = 134 K. \textbf{(a)} Schematic cross-section of the planet; layer thicknesses are to scale. The atmosphere corresponds to pressures below 1000 bar. (b) Pressure-temperature (P-T) profile of the atmosphere (light blue) and interior (blue and teal). (c) Transmission spectrum for this model, with the three observed transmission datasets shown for comparison. d) Emission spectrum for this model, compared against the four observed emission datasets. (e) Interior metal mass fraction profile ($Z$; black) and H/He mass fraction as functions of normalized radius. The atmospheric region ($P<$ 1000 bar) is indicated in light blue. (f) Atmospheric mass mixing ratios of the molecular species comprising the metals ($P<$ 1000 bar). The region highlighted in orange indicates the pressure range probed by emission spectra, as calculated from the contribution function.}
         \label{fig:master_plot}
\end{figure*}

% Last results section. Summary of joint interior + atmosphere retrievals. These are 4 retrievals: transmission with and without the age; emission with and without the age. Equilibrium chemistry is assumed across these because (1) it is expected based on WASP-80 b equilibrium temperature 

The joint interior-atmosphere retrievals presented in this section incorporate into their likelihood function both spectra and bulk observables (mass, radius, age), following Sect. \ref{sec:log-likelihood}. In addition, the forward model consists of the coupled interior-atmosphere model described in Sections \ref{subsec:gastli_joint_mode}, \ref{sec:petitradtrans} and \ref{sec:coupling_joint}. The suite of joint retrievals (JR; see Table \ref{tab:joint_intatm_summary}) consists of six retrievals: two with transmission spectra (JR1, JR2), two with emission spectra (JR3, JR4) and two that combine both geometries (JR5, JR6). We assume equilibrium chemistry across the suite of joint retrievals because the transmission-only retrievals  show a preference for the equilibrium chemistry model (see Sect. \ref{sec:transmission_only}). Additionally, we explored JR1 and JR5 as free chemistry retrievals. Table \ref{tab:joint_intatm_summary} shows a summary of the compositional parameter estimates (CMF, $Z_{\rm planet}$, log(M/H), C/O) from joint retrievals. Furthermore, Fig. \ref{fig:master_plot} shows a summary schematic of JR6, which we adopt as our fiducial joint retrieval (see Sect. \ref{sec:clouds_joint}).

%and discuss their results in the context of the tension between the atmospheric metallicity and the density presented in Sect. \ref{sec:interior_only_retrievals} at the end of this section.

%In the retrievals that include the transmission spectra -- all except for JR3 and JR4 -- the planet radius is not included as a bulk observable in $\mathbf{d_{\rm bulk}}$ because this information is already contained in the transmission spectrum. 

%Consequently, the difference of JR1 and JR5 with respect to JR2 and JR6, respectively, is that the latter include the age in addition to the planet mass. 

%Finally, we assumed equilibrium chemistry across the complete suite, as this is expected at the equilibrium temperature of WASP-80 b. Nonetheless,

% Add a summary table of the compositional parameters, age and cloud parameters + 

\begin{table*}[h]
\caption{\label{tab:joint_intatm_summary} Summary of the retrieved model parameters in the joint interior-atmosphere retrievals.}
\resizebox{\textwidth}{!}{%
\centering
\begin{tabular}{clcccccccc}
\hline
Retrieval name & $\mathbf{d_{\rm bulk}}$ & $\mathbf{d_{\rm spectrum}}$ & log(M/H) & C/O & CMF & $Z_{\rm env}$ & $Z_{\rm planet}$ & $T_{\rm int}$ [K] & log($P_{\rm base}$) [bar] \\ \hline \hline
JR1 & $\{ M_{\rm pl} \}$ & T & $-0.48^{+0.35}_{-0.30}$ &  0.35$^{+0.16}_{-0.14}$ & 0.14$^{+0.07}_{-0.09}$ & 0.01$\pm$0.01 & 0.14$^{+0.07}_{-0.08}$ & 987$^{+91}_{-95}$ & 0.50$^{+1.66}_{-1.44}$ \\
JR1$^{\ast}$ (free chem.) & $\{ M_{\rm pl} \}$ & T & 0.38$^{+0.49}_{-0.51}$ & 0.06$^{+0.02}_{-0.04}$ & 0.12$^{+0.08}_{-0.07}$ & 0.05$\pm$0.04 & 0.17$^{+0.07}_{-0.08}$ & 1173$^{+268}_{-293}$ & $-1.91^{+1.53}_{-1.14}$ \\
JR2 & $\{ M_{\rm pl}$, Age$\}$ & T & $-0.53^{+0.35}_{-0.28}$ & 0.43$^{+0.13}_{-0.16}$ & 0.02$^{+0.02}_{-0.01}$ & (6.8$\pm$5.2)$\times 10^{-3}$ & 0.03$\pm$0.02 & 75$^{+25}_{-38}$ & 0.88$^{+1.78}_{-1.66}$ \\
JR3 & $\{ M_{\rm pl}, R_{\rm pl}\}$ & E & 0.69$^{+0.30}_{-0.27}$ & 0.18$^{+0.09}_{-0.06}$ & 0.25$^{+0.23}_{-0.16}$ & 0.14$\pm0.05$ & 0.37$\pm0.16$ & 300$^{+123}_{-106}$ & - \\
JR4 & $\{ M_{\rm pl}, R_{\rm pl}$, Age$\}$ & E & 0.46$^{+0.20}_{-0.17}$ & 0.15$^{+0.06}_{-0.03}$ & 0.04$^{+0.04}_{-0.03}$ & 0.09$\pm0.02$ & 0.14$\pm0.03$ & 115$^{+13}_{-12}$ & - \\
JR5 & $\{ M_{\rm pl} \}$ & T, E & 0.20$^{+0.07}_{-0.08}$ & 0.11$\pm$0.01 & 0.48$^{+0.03}_{-0.04}$ & 0.08$\pm$0.01 & 0.52$\pm$0.03 & 416$^{+30}_{-24}$ & 0.21$^{+1.56}_{-1.39}$ \\
JR5$^{\ast}$ (free chem.) & $\{ M_{\rm pl} \}$ & T, E & 1.00$^{+0.26}_{-0.28}$ & 0.24$\pm$0.09 & 0.18$^{+0.12}_{-0.10}$ & 0.12$\pm$0.06 & 0.28$\pm$0.11 & 257$^{+71}_{-68}$ & $-0.29^{+2.09}_{-2.30}$ \\
JR6 & $\{ M_{\rm pl}$, Age$\}$ & T, E & 0.44$^{+0.12}_{-0.11}$ & 0.12$^{+0.03}_{-0.02}$ & 0.02$^{+0.02}_{-0.01}$ & 0.10$\pm$0.02 & 0.12$\pm$0.02 & 134$^{+12}_{-9}$ & 0.09$^{+2.60}_{-4.48}$ \\ \hline
\end{tabular}%
}
\tablefoot{These include the atmospheric composition parameters, log(M/H) and C/O; the planet bulk compositional parameters, CMF, $Z_{\rm env}$ and $Z_{\rm planet}$; the internal (or intrinsic) temperature $T_{\rm int}$, and one of the atmospheric cloud parameters, log($P_{\rm base}$). T = Transmission, E = Emission. $^{\ast}$ All joint retrievals consider equilibrium chemistry except for this one, which is similar to JR1 but assumes free chemistry.}
\end{table*}

% comparison between joint retrievals: (1) atm. composition

%\subsection{Comparison to interior and atmospheric retrievals}
\subsection{Bulk and atmospheric composition}
% "checks"

\begin{figure}[h]
   \centering
   \includegraphics[width=\columnwidth]{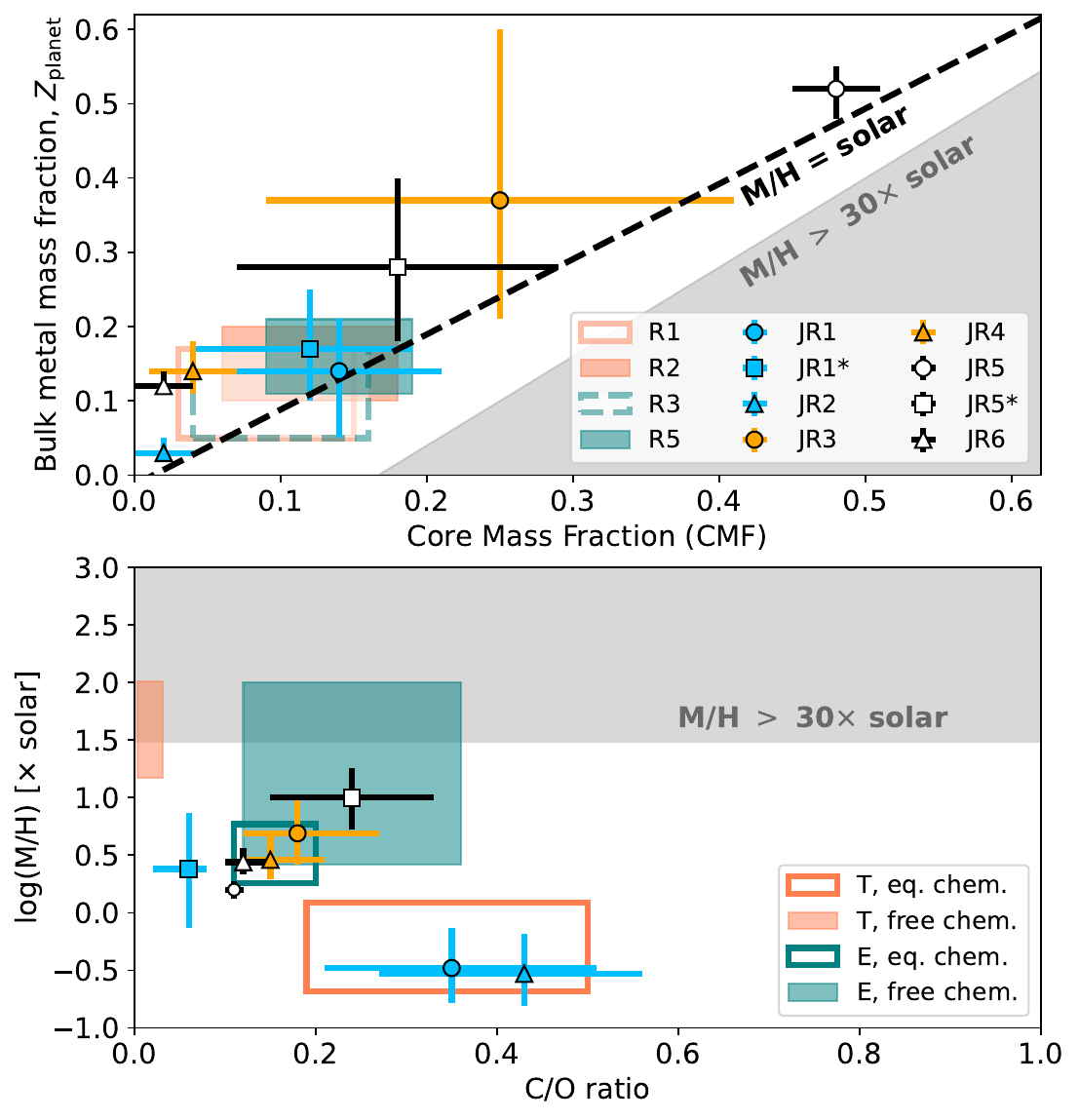}
      \caption{Top panel: Core and bulk metal mass fraction 1-$\sigma$ estimates obtained by our joint retrievals (JR1-JR6). Joint retrievals that take into account the age have triangle markers. Interior-only retrievals (R1-R3 and R5) are shown as colored boxes for comparison. Bottom panel: Atmospheric metallicity and C/O ratio estimates from our suite of joint retrievals. Spectra-only estimates are displayed as colored boxes for comparison.}
         \label{fig:joint_summary}
\end{figure}

The compositional parameters derived in the joint retrievals are compared in Fig. \ref{fig:joint_summary} with the interior-only retrievals (top panel: CMF and $Z{\rm planet}$) and the atmosphere spectrum retrievals (bottom panel: C/O and M/H). As the joint retrievals combine the observations used in the independent interior and atmosphere retrievals simultaneously, their estimates should be consistent with these.  With the exception of JR5, joint retrievals agree within 2$\sigma$ with the interior-only retrievals, as shown by their overlap in CMF-$Z_{\rm planet}$ parameter space (Fig. \ref{fig:joint_summary}, top panel). JR5, which incorporate emission spectra without the age, tends to overestimate the CMF. We discuss the cause of this high-CMF estimate in Sect. \ref{sec:joint_pts}.

The bulk metal and core mass fraction estimates align along the solar envelope line (dashed black line in Fig. \ref{fig:joint_summary}). We compare the envelope composition estimates in the bottom panel. The joint retrievals that do not include emission spectra (JR1, JR2) retrieve a sub-solar atmospheric metallicity and a solar C/O ratio, being consistent with the fiducial transmission-only retrieval. JR3-JR6, which include emission data with bulk properties and/or the transmission spectrum, obtain atmospheric metallicities and C/O ratios in excellent agreement with the emission-only retrievals (Fig. \ref{fig:joint_summary}, bottom panel). These show that WASP-80 b has a sub-solar C/O$\sim$0.10-0.20 and [M/H] = 1-20$\times$ solar. JR1 and JR2 (transmission) are compatible with solar C/O values, in contrast to the sub-solar C/O in the other joint retrievals. 

Overall, we demonstrate that the joint retrievals that include emission spectra are consistent with the emission-only retrievals, while JR1 and JR2 are consistent with the transmission-only retrievals. We observe that the bulk composition derived in the joint retrievals are also consistent with the traditional interior-only retrievals, with the exception of JR5.

\subsection{The M/H discrepancy and degeneracies in joint retrievals}

We showed in Sect. \ref{sec:transmission_only} that the atmospheric metallicity derived from the free chemistry, transmission-only retrieval is strongly super-solar, and incompatible with the interior-only retrievals that assume secular cooling. To explore this tension, we modified our joint retrieval JR1, which takes into account the transmission spectrum together with the mass as observables, as a free chemistry retrieval. This free chemistry retrieval, JR1$^{\ast}$, tends to obtain a lower C/O ratio and higher atmospheric metallicity than its equilibrium counterpart. This decrease in C/O and increase in M/H is consistent with the behaviour we observed in the transmission-only retrievals, as the free chemistry retrieval exhibits a preference for a water-rich atmosphere and high-altitude clouds. In contrast to the transmission-only, free retrieval, the atmospheric metallicity has an upper 1$\sigma$ limit of log(M/H) = 0.89 , being below the 1$\sigma$-upper limit derived by the interior-only analysis of 10 $\times$ solar. This is caused by the coupling between the reference radius used to generate the transmission spectrum and the interior model. Thus, the coupling between temperature, chemistry (metal content) and radius in joint retrievals effectively demonstrates that there is a degeneracy between the envelope composition (M/H and C/O) and chemistry.

\subsection{Clouds}
\label{sec:clouds_joint}

We obtain very similar estimates on the cloud parameters between each joint retrieval, and these and the transmission-only retrievals in Sect. \ref{sec:transmission_only}. For reference, we have included the cloud base pressure in Table \ref{tab:joint_intatm_summary}. Most estimates are consistent with a cloud base located at $P> 0.1$ bar at 1$\sigma$. The most data-complete joint retrieval, JR6, suggests that clouds are present at high altitudes $ P \sim 10^{-4}$ bar within 1$\sigma$. 

We saw in Sect. \ref{sec:transmission_only} that there is a degeneracy between the opacity of carbon- and oxygen-bearing molecules and cloud opacities in spectra. From the least degenerate joint retrieval (JR6), we can conclude that a sub-solar C/O ratio together with an atmospheric metallicity [M/H] = 1-10$\times$ solar and high altitude clouds is the most likely scenario for WASP-80 b, as this is compatible with all spectral data sets and the planet bulk density simultaneously, under the assumption of equilibrium chemistry.

\subsection{Thermal structure}
\label{sec:joint_pts}

\begin{figure*}[h]
   \centering
  \includegraphics[width=\textwidth]{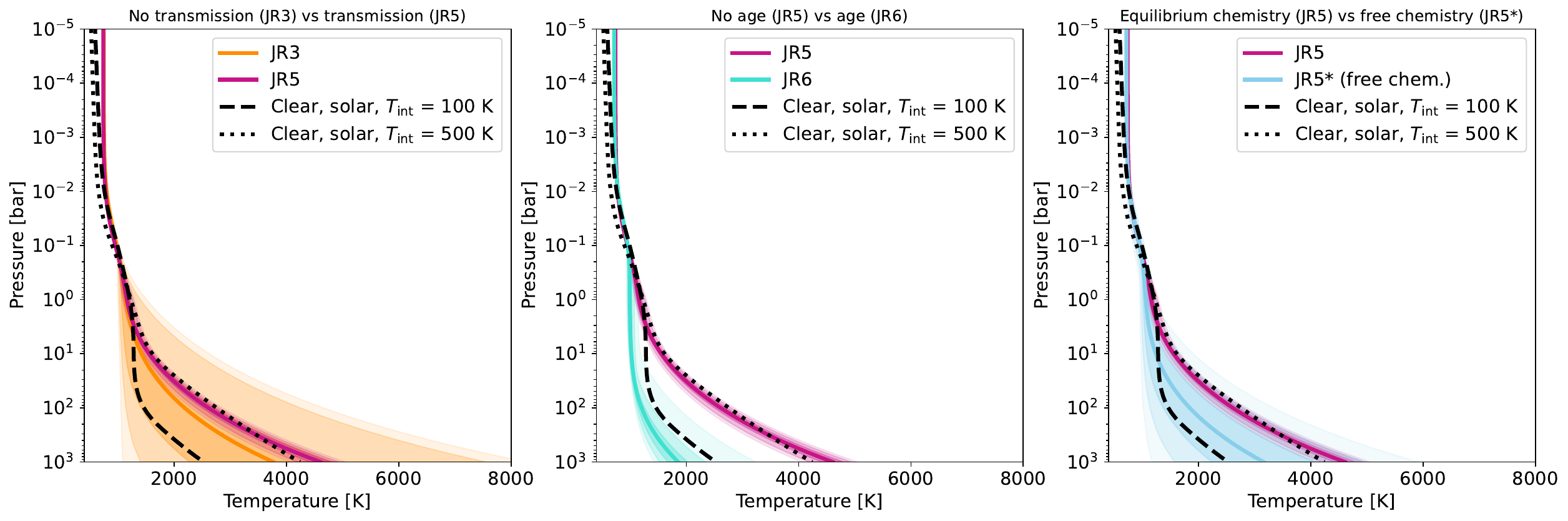}
      \caption{Pressure-Temperature (PT) profiles of WASP-80 b as constrained by four of our joint retrievals. The shaded regions indicate the 1, 2, and 3$\sigma$ regions of the retrieval, while solid lines correspond to the mean. Black lines highlight self-consistent models from \cite{Molliere17,Acuna24_gastli_science_paper} of a clear atmosphere with solar composition at $T_{\rm int}$ = 100 (dashed) and 500 K (dotted) for comparison. \textbf{Left}: The difference between JR3 and JR5 shows the effect of adding the transmission spectrum to JR3, which incorporates emission spectra plus the mass and radius. The transmission spectrum reduces significantly the uncertainties in thermal structure. \textbf{Center}: Comparing JR6 and JR5 allows us to study the effect of including the age to the retrieval. JR6 is the most-data complete of the joint retrievals, since it incorporates both transmission and emission, plus the mass and age. If the age is removed (JR5), the temperature in the deep atmosphere increases. \textbf{Right}: The comparison between JR5 and JR5$^{\ast}$ illustrates the effect of assuming equilibrium chemistry in the joint retrieval. If we relax the assumption of equilibrium chemistry in JR5 by changing to a free chemistry model (JR5$^{\ast}$), the temperature decreases in the deep interior.}
         \label{fig:ptprofile_joint}
\end{figure*}

% (2) thermal structure IN GENERAL
We show in Fig. \ref{fig:ptprofile_joint} the thermal structure profiles constrained by all our joint retrievals (except for JR4) that incorporate emission spectra. These profiles agree well with those derived by the emission-only retrievals (see Sect. \ref{sec:emission_only}) at pressures $P < 1$ bar. As demonstrated in Fig. \ref{fig:PT_profile_emission} by the contribution function, these are the pressures probed by the emission spectrum. At higher pressures ($P = 1-1000$ bar), the constrained thermal structure is very dependent on the assumptions and data included in the retrieval. In the following, we discuss how these impact the temperature in the deep envelope and the bulk metal mass fraction.

% age vs no age
Including the age in the likelihood improves the constraints on the internal temperature. This can be seen in the central panel of Fig. \ref{fig:ptprofile_joint}, and by comparing the internal temperatures in Table \ref{tab:joint_intatm_summary}. When the age is not taken into account, the internal temperatures have wide uncertainties ($\Delta T_{\rm int} \sim 100-200$ K), and their mean values tend to be high ($T_{\rm int} > 200$ K). As WASP-80 is a moderately old system (0.99 Gyr), the age's effect in the retrievals is to decrease the internal temperature to $T_{\rm int} =$50-150 K, with uncertainties as low as $\Delta T_{\rm int} = $ 12 K in the least degenerate case, JR6. This range of internal temperatures is consistent with the interior-only retrievals that account for the age (R1, R5-R6) and the interior forward models (see Fig. \ref{fig:interior_forward_nocore}). 

Independently of the geometry considered, adding the age to the likelihood reduces the mean value of the retrieved bulk metal mass fraction and the CMF (see Fig. \ref{fig:joint_summary}, top panel). As discussed above, the age of WASP-80 b decreases the internal temperature, which entails a lower temperature at 1000 bar. Since the interior temperature follows an adiabatic profile, a lower $T_{\rm surf}$ correlates to a colder interior. A colder interior adiabat increases the density of the envelope, which lowers the retrieved CMF and $Z_{\rm planet}$.

% effect of transmission spectra  
The inclusion of transmission spectra in the joint retrievals tends to increase the inferred temperature in the deep envelope and reduce its uncertainties. This effect can be seen when comparing $T_{\rm int}$ in JR1 and JR3 (Table \ref{tab:joint_intatm_summary}), and in the left panel of Fig. \ref{fig:ptprofile_joint}. When comparing JR1 (transmission without emission) and JR3 (emission without transmission), we find that JR3 yields lower internal temperatures than JR1. JR3 is more sensitive to the temperature at 1 bar -- $T(P =$ 1 bar) = 700-900 K (see also Sect. \ref{sec:emission_only}) --, favoring lower $T_{\rm int}$. Similarly, when both transmission and emission are considered (JR5), the temperature increases compared to JR3 (no transmission, see left panel of Fig. \ref{fig:ptprofile_joint}). In Sect. \ref{sec:transmission_only}, we noted that the fiducial transmission retrieval tends to retrieve low C/O ratios. We discussed that this could be due to the assumption of equilibrium chemistry, where low C/O ratios and high deep atmospheric temperatures are required to increase the abundance of H$_{2}$O relative to CH$_{4}$. Including the transmission spectrum along with the emission spectrum in a joint retrieval further constrains the abundances of H$_{2}$O and CH$_{4}$, which drives model in JR5 toward a higher internal temperature ($T_{\rm int}$ = 500 K). This high deep atmospheric temperature increases the planet radius, yielding a core and bulk metal mass fraction in JR5 that is the highest of all joint retrievals (see Fig. \ref{fig:joint_summary}, top panel).

% effect of equilibrium chemistry
Finally, we explore the effect of assuming equilibrium chemistry on the retrieved thermal structure. We compare the P-T profile of JR5 (emission plus transmission, without age) between equilibrium and free chemistry (JR5$^{\ast}$). As discussed above, the equilibrium chemistry model requires high temperatures to reproduce the abundance of H$_{2}$O observed in the spectra. Thus, relaxing the equilibrium assumption in the free chemistry retrieval leads to a decrease in the temperature at high pressures, as shown in the left panel of Fig. \ref{fig:ptprofile_joint}. The thermal structure of JR5$^{\ast}$ is compatible with a self-consistent, clear solar model with $T_{\rm int} = 100$ K at 1$\sigma$, which is the internal temperature predicted by the age of WASP-80 b in the absence of extra heating. The decrease in deep atmospheric temperature between JR5 and JR5$^{\ast}$ has a significant impact on the retrieved CMF. In Fig. \ref{fig:joint_summary} (top panel), we can see that the CMF decreases from 0.50 (JR5) to $\sim$0.20 (JR5$^{\ast}$). Since JR5$^{\ast}$ (free chemistry) is more consistent with the other retrievals than JR5, it represents a more conservative scenario than JR5. This is especially the case if WASP-80 b's internal temperature is decoupled from its age due to additional heating sources. Thus, we adopt JR5$^{\ast}$ as our fiducial retrieval, together with JR6. The latter represents a scenario without extra heating sources where age and $T_{\rm int}$ are coupled.

\subsection{Precision in bulk metal mass fraction, $Z_{\rm planet}$}

In the following, we compare the CMF and $Z_{\rm planet}$ estimates shown in Table \ref{tab:joint_intatm_summary} to discuss how the different observables and data sets can improve the precision in bulk metal mass fraction. For this comparison, our reference uncertainty of the bulk metal mass fraction is $\Delta Z_{\rm planet}$ = 0.05, which is the uncertainty of the most-data complete of the traditional, interior-only retrievals, R5-R6. 

We note that all three joint retrievals that incorporate the age -- JR2, JR4, and JR6 -- have bulk metal mass fraction uncertainties of $\Delta Z_{\rm planet}$ = 0.02-0.03. This means that a joint retrieval together with age reduces the uncertainties from 0.05 to 0.02 -- a factor of 1/2. However, in the absence of an age estimate, a joint retrieval does not guarantee an improvement of the retrieved bulk metal mass fraction relative to the traditional, interior-only retrievals. For example, R3 has a $\Delta Z_{\rm planet, \ R3}$ = 0.05-0.07, while its joint counterpart in transmission (JR1) shows a similar uncertainty $\Delta Z_{\rm planet, \ JR1}$ = 0.07. Furthermore, the joint counterpart in emission (without transmission), JR3, obtains a larger uncertainty of $\Delta Z_{\rm planet, \ JR3}$ = 0.16. This is due to the degeneracy between temperature in the deep atmosphere ($T_{\rm surf}$) and CMF, which is mostly solved by the age.

\section{Discussion}
\label{sec:discussion}

\cite{Wilkinson24} conducted a similar analysis using the MCMC method and a coupled interior-atmosphere model on WASP-39 b's transmission spectrum. There are three key differences between their analysis and this work: first, we include a condensate opacity source to simulate the effect of clouds; second, we do not weight the importance of the bulk density observables i.e. mass, radius and age; third, our retrievals incorporate data in both the near-infrared (nIR) and optical. We examine how cloud treatment, weighted likelihood, and the spectral wavelength coverage affect the inferred atmospheric and bulk compositions in atmosphere-only and joint retrievals in Sects. \ref{sec:cloud_wilkinson}, \ref{sec:likelihood_wilkinson} and \ref{sec:disc_previous}, respectively.

Furthermore, we discuss the possible presence of external heating sources and disequilibrium chemistry in WASP-80 b in Sects. \ref{sec:hightemp_disc} and \ref{sec:disequilibrium_disc}. Finally, we conclude by outlining our model uncertainties and caveats (Sect. \ref{sec:caveats}) and by discussing the implications of WASP-80 b's atmospheric and bulk composition for its formation pathways (Sect. \ref{sec:formation_disc}).

%In the following, we discuss how each of these modeling choices can impact the inferences on atmospheric and bulk compositions.

\subsection{Sensitivity to cloud treatment} \label{sec:cloud_wilkinson}

% (1) clouds need to be taken into account, Wilkinson+ results on WASP-39 b are likely biased. 

The omission of cloud modelling in joint retrievals with transmission spectra may lead to biases in the inference of atmospheric composition and bulk metal mass fraction. In traditional, transmission-only retrievals, the atmospheric metallicity and the cloud top pressure ($P_{\rm base}$) are often correlated \citep{Benneke12,Welbanks16_degeneracies,Fisher18}. To study the impact of not including clouds in our modelling, we run our JR1 retrieval without clouds. We obtain CMF$_{\rm JR1, \ clear}$ = 0.09$\pm$0.05, log(M/H)$_{\rm JR1, \ clear}$ = $-0.92^{+0.11}_{-0.06}$ and C/O$_{\rm JR1, \ clear}$ = 0.12$^{+0.03}_{-0.01}$. If we compare these to the fiducial JR1 retrieval, we find that the uncertainties in atmospheric metallicity and C/O ratio are reduced by a factor of 3 and 5, respectively. This bias arises from the reduced parameter space caused by eliminating the degeneracy between chemical abundance and cloud pressure. The CMF mean is decreased from 0.14 to 0.09, while their uncertainties are reduced by a factor of 2. We conclude that not accounting for clouds in joint interior-atmosphere retrievals biases not only the inference of atmospheric composition, but also the inferred CMF and $Z_{\rm planet}$. These biases include lower C/O ratio and CMF mean values, and significantly reduced uncertainties for all compositional parameters, by factors of 2-4.

% The second degeneracy concerns the atmospheric abundances and the location of the cloud, $P_{\rm base}$, because a gray cloud deck has a similar effect on the transmission spectrum as an optically thick atmosphere. This is implicitly included in our cloud model with the parameter $\kappa_{0}$ because it regulates the opacity of the cloud layer: $\kappa_{0}$ = 0 corresponds to a clear atmosphere, while $\kappa_{0} >> 1$ represents a fully cloudy envelope. Consequently, clear atmosphere retrievals may introduce biases in the estimates of C/O and log(M/H) in joint retrievals. To test this, 
% (2) About using a scale factor for the interior observables: we recommend not to do that because it is equivalent to decreasing the uncertainties of the mass, age.

\subsection{Effect of weighted likelihood} \label{sec:likelihood_wilkinson}

The second key difference from previous work is that we do not scale the terms associated with the bulk density observables in Eq. \ref{eq:likelihood_joint} ($\mathbf{d_{\rm bulk}}$) by a weighting factor ($w_{j}$). \cite{Wilkinson24} fix $w_{j}$ to a constant value that equalizes the importance of these observables with that of the spectrum. Under this assumption, the weighting factor must be equal to the total number of points in the spectrum. A value $w_{j} > 1$ is equivalent to reducing the uncertainties of the observables, because the factor $ w_{j} / \sigma^{2}_{\rm bulk, \ j} $ in Eq. \ref{eq:likelihood_joint} can also be written as $ 1 / \left( \sigma_{\rm bulk, \ j} / \sqrt{w_{j}} \right)^{2}   $. In volatile-rich planets, reducing mass uncertainties does not improve the precision of inferred bulk composition since the envelope mass contributes a negligible fraction to the total mass \citep{Otegi20}. However, a reduction in radius and age uncertainties significantly improves the precision of the inferred CMF and $Z_{\rm planet}$ \citep{Muller23}. Consequently, using a weighting factor to give more importance to the bulk density observables relative to the transmission spectrum leads to an underestimation of the uncertainties of core and bulk metal mass fraction.

%We therefore advise against using weights to give more importance to the bulk density observables with respect to the transmission spectrum, especially if radius and age are taken into account.

% add figure with histogram to show that not using the weights can reproduce perfectly the observed distributions of mass, radius and age?

\subsection{Effect of wavelength coverage in atmospheric spectra} \label{sec:disc_previous}

In contrast to WASP-39 b's joint retrievals by \cite{Wilkinson24}, our analysis incorporates both nIR and optical transmission spectra. Transmission spectra retrievals exhibit a normalization degeneracy, in which the molecular abundances relative to H are degenerate with the absolute normalization of the spectrum. The normalization depends on the reference pressure, $P_{\rm ref}$. This degeneracy can be broken if the spectrum is sensitive to $P_{\rm ref}$ through pressure broadening or collision-induced absorption (CIA). The transmission spectrum contains information about these processes when it spans a wide wavelength range in the nIR, such as that provided by the combined HST and JWST datasets used in our analysis \citep{Fisher18}. Additionally, optical data further reduce degeneracies between the chemical abundances and the cloud top pressure \citep{Fairman24}. The optical wavelength range provides valuable information about condensates in transmission retrievals \citep{Welbanks16_degeneracies}. 

% discuss more clouds from Morel+ and possible effects on inference

We further compare our results to those of \cite{Morel25}, who use reflected light data in traditional, atmosphere-only retrievals. They provide and analyse a dataset of WASP-80 b emission spectra between 0.6 and 3 $\mu$m, measured with JWST's NIRISS/SOSS. Their retrievals rule out log(M/H) $> 2$ and C/O $> 0.6$, which is in good agreement with our joint retrievals. Their self-consistent cloud modelling concludes that the NIRISS/SOSS data are compatible with cloud species such as chromium (Cr[s]), sodium sulfide (Na$_{2}$S), potassium chloride (KCl) and Zinc sulfide (ZnS). \cite{Morel25} also consider the presence of tholins in the atmosphere of WASP-80 b. In this scenario, they estimate a low tholin production rate ($< 10^{-11.5} {\rm \ g \ cm^{-2} \ s^{-1}}$). Tholins tend to cool the envelope due to a high UV extinction, as observed in the sub-Neptune GJ1214 b \citep{Nixon25} and as predicted in cold, directly imaged planets \citep{Morley15}. Our fiducial retrieval (JR6) is compatible within uncertainties with a cool atmosphere (see Fig. \ref{fig:ptprofile_joint}, central panel). Additionally, our atmosphere-only and joint retrievals further support the presence of condensates in transmission, (see Sect. \ref{sec:transmission_only} and \ref{sec:joint_retrievals}), being consistent with the scenario proposed by \cite{Morel25}.

%This wavelength range in emission does not have significant molecular features, thus including it in our joint retrievals would not further constrain the atmospheric compositional parameters. 

Our analytical parameterization and priors of the thermal structure are conservative and flexible enough to accommodate a wide range of Bond albedos (see Figs. \ref{fig:PT_profile_emission} and \ref{fig:ptprofile_joint}). \cite{Morel25} constrain the Bond albedo to $A_{B} = 0.15-0.40$ at 1$\sigma$. This interval could potentially further constrain the thermal structure in our fiducial joint retrieval. Thus, the bulk composition obtained from joint retrievals that include reflected light would lie within the confidence intervals reported in Sect. \ref{sec:joint_retrievals}. Although the analysis of reflected light data is beyond the scope of this work, our future efforts will focus on incorporating the NIRISS/SOSS dataset into our joint interior–atmosphere framework. Such an analysis could further constrain the retrieved thermal structure profile if the reflected light data is combined with a grid of self-consistent cloudy PT profiles \citep{charnay2018,lacy_burrows,diamondback_clouds} and condensation curves \citep{virga}. These condensation curves can determine the location and composition of the condensate layer instead of assuming free cloud parameters in our retrievals. 

%When clouds are included, self-consistent thermal structure models have higher temperatures at the bottom of the atmosphere induced by the cloud opacity. This delays the thermal evolution of the interior, producing a higher internal temperature for the same age \citep{Linder19,Poser19,Poser24}. In contrast,  Including reflected light in a joint retrieval with self-consistent cloud models could potentially increase (decrease) the mean CMF estimate, as a warmer (colder) interior requires more (less) metals to fit the same radius. Nonetheless, 

\subsection{Non-adiabatic interiors and heating mechanisms} \label{sec:hightemp_disc}

% (2) similar to Acuña et al. 2024, bulk metal content may change if a self-consistent dilute core is considered instead of a compact core. Nonetheless, this approach is still useful for homogeneous analyses in the context of planet formation mechanisms. 

A key assumption in our analysis is that the deep interior of WASP-80 b is adiabatic. Several mechanisms can raise the temperature above that of an adiabatic interior, such as compositional gradients (i.e., an inhomogeneous interior), tidal heating or Ohmic dissipation. Super-adiabatic interior models could accommodate more metals in the bulk composition of WASP-80 b. For instance, in Sect. \ref{sec:interior_only_retrievals}, we noted that WASP-80 b's density and age could be compatible with a super-solar atmospheric metallicity of M/H $> 30 \times$ solar, if its interior is warmer than our fiducial, interior-only model. In this section, we assess the extent to which these processes could affect the inference of WASP-80 b's bulk composition.

% non-homegeneity

Compositional gradients induce hotter, non-adiabatic interior temperature profiles \citep{Vazan18}, resulting in larger planetary radii than homogeneous envelope models. This raises the question of whether WASP-80 b could have an inhomogeneous interior, thereby affecting the bulk metal mass fraction inferred by our homogeneous model. Inhomogeneous envelopes have been confirmed in Jupiter and Saturn \citep{Wahl17,Nettelmann21,Mankovich21,Miguel22}, and are a natural outcome of planet formation in both the Solar System \citep{Helled_Stevenson17,Lozovsky17,ValletaHelled20,Stevenson22} and exoplanets \citep{bodenheimer18,Ormel21,Brouwers21}.  \cite{Knierim24} conducted a parameter study to explore the planet masses and primordial entropies ($S_{0}$) at which compositional gradients are expected. They find that at WASP-80 b's mass, fast mixing in the deep envelope is efficient enough to erase any compositional gradients from formation, regardless of the initial entropy (see their figures 4 and 6). Thus, a planet with WASP-80 b's mass and age would be be expected to have a homogeneous interior.

Moreover, helium rain could also affect the inference of the bulk metal mass fraction in extrasolar gas giants. \cite{Knierim25} included the effect of helium rain in interior models with equilibrium temperatures relevant to that of WASP-80 b ($T_{\rm eq} = 500 $ K). Helium rain is predicted to have a negligible effect on the planetary radius at WASP-80 b's equilibrium temperature and mass, with $\Delta R_{\rm planet} < 0.01 \ R_{\rm Jup}$ \citep[see figure B.1 in][]{Knierim25}. Consequently, the omission of compositional gradients and helium rain in our modelling is unlikely to affect our inferred atmospheric and bulk composition of WASP-80 b.

%Our interior structure model assumes that the envelope metallicity is constant with radius, $Z(r) = Z_{\rm env}$, and that the core is compact.

%In addition, thermal cooling is delayed if compositional gradients are present as a consequence of mixing being suppressed and convection being inhibited \cite{Knierim24}.

%Following their figures 4 and 6,  would have an atmosphere-to-bulk metal mass fraction ratio of $Z_{\rm atm}/Z_{\rm bulk} = 0.9-1$, which is indicative of a high-degree of homogeneity. This estimate corresponds to $S_{0} = 9 \ k_{B} m_{H}$, which is representative of a cold-start formation scenario \citep{Marley07,SB12}. Higher primordial entropies $S_{0} = 9.5-11 \ k_{B} m_{H}$ lead to an increase of $Z_{\rm atm}/Z_{\rm bulk}$ for a constant mass and age. Consequently, WASP-80 b is likely to have a homogeneous envelope independently of its primordial entropy. 

%Furthermore,  For planets whose masses are $M < 1 \ M_{\rm Jup}$, the increase in interior temperature caused by helium rain has a negligible effect on radius with $\Delta R_{\rm planet} < 0.01 \ R_{\rm Jup}$ (see their figure B.1). Thus, with a mass of 0.55 $M_{\rm Jup}$, WASP-80 b's bulk metal mass fraction estimated by assuming a homogeneous envelope is not expected to be impacted by compositional gradients or helium rain. 

% tidal heating

Warm gas giants can experience internal heating due to external sources \citep{Millholland20,Vissapragada24}. This has been confirmed in WASP-107 b via disequilibrium chemistry, which shows its internal temperature rising from $T_{\rm int} \sim$100 K to $T_{\rm int} >$ 460 K \citep{Sing_24_WASP107b,Welbanks_WASP107b}. It is currently under debate whether WASP-107 b is inflated due to tidal heating or Ohmic dissipation \citep{Batygin25}. \cite{Triaud15} estimate an eccentricity for WASP-80 b of $e = 0.002^{+0.010}_{-0.002}$. We use the tidal heating model in \cite{Leconte2010} to estimate the internal temperature induced by tidal heating in WASP-80 b, assuming its mean eccentricity. For $Q' > 2.7 \times 10^{5}$ - Jupiter's minimum value - WASP-80 b's internal temperature is $T_{\rm int} = 200$ K for a Love number and time lag factor of $ k_{2} \times \Delta t$ = 0.02 s. Since internal temperature increases for lower tidal dissipation factor values, we adopt 200 K as a conservative upper limit on the internal temperature induced by tidal heating. Thus, we conclude that tidal heating is unlikely to be inflating the interior of WASP-80 b.

Ohmic dissipation may also play an important role in the interior structure of warm Jupiters. This mechanism results from the interaction between the planetary magnetic field and atmospheric winds. It has been studied in detail for hot Jupiters and is a leading explanation for radius inflation at $T_{\rm eq} > 1000$ K \citep{Perna10,Batygin10,Thorngren18}. More recently, it has also been proposed as a source of inflation for smaller, cooler planets, including WASP-107b and sub-Neptunes \cite{Batygin25,PuValencia17}. WASP-80 b lies between these two regimes in both mass and equilibrium temperature, so Ohmic dissipation may be relevant. Furthermore, for a fixed mass and equilibrium temperature, the extent of heating depends on wind depth, wind velocity, and magnetic field intensity \citep{Ginzburg16,Komacek17,Knierim22}. Magnetic fields and winds can be probed using ground-based, high-resolution spectroscopy \citep{Ehrenreich20,Keles21} and spectropolarimetry \citep{spectropolarimetry}. Ultra-hot and hot Jupiters are currently the only class of planets accessible to these techniques.

Although wind and magnetic field properties remain unconstrained for warm gas giants, the effect of magnetic fields in this temperature and mass regime can be suggested via other observational methods. A non-detection of the metastable-helium triplet (1083 nm) signature in transit was reported for WASP-80 b, despite being expected by hydrodynamical escape predictions due to the low planetary density \citep{Fossati22,shreyas_22}. One hypothesis to explain this lack of helium absorption is that planetary magnetic fields can disrupt the ideal assumption of the neutral gas forming a symmetric shell or extended tail around the planet disk. Instead, a strong magnetic field can channel ionized outflows along the open field lines near the poles. This scenario is termed magnetic confinement, as the magnetic pressure is high enough to prevent the outflow to expand isotropically \citep{Adams11,Trammell11}. In addition, an intense planetary magnetic field can also act as a bottleneck to the flow at the Alfvén surface, suppressing the mass flux escaping the planet due to photoevaporation \citep{Owen14}. \cite{Yadav_Thorngren17} estimate the magnetic field strength for the hot Jupiter sub-population based on their observed inflation and a luminosity-dynamo scaling model. Jupiter's field strength corresponds to 10 G \citep{Moore17}, while hot Jupiter-mass planets can have field strengths ranging from 10 to 100 G. Their model suggests that at constant mass, a higher equilibrium temperature increases the field strength by a factor of 10. This behaviour could be conserved in Saturn-mass planets, closer to WASP-80 b's mass. Saturn's magnetic field strength is 0.2 G at $T_{\rm eq}  < 100$ K \citep{Russell93,Belenkaya06}. For WASP-80 b at $T_{\rm eq}  = 825$ K, the field strength could be has high as 2 G. This field strength would be sufficient to magnetically confine or suppress the outflow at the Alfvén surface \citep{Fossati23}. Thus, the scenario we propose in our analysis where WASP-80 b has a super-solar envelope and is inflated by a heating mechanism could be explained, together with independent He triple transit observations -- by Ohmic dissipation.

%\textcolor{red}{Add topic sentence at the end of this paragraph (magnetic fields can explain these observations and our free chemistry scenario) + references. Split in two paragraphs if too long.}

\subsection{Effects of disequilibrium chemistry} \label{sec:disequilibrium_disc}

%Heating due to Ohmic dissipation would drive the atmosphere of WASP-80 b out of chemical equilibrium.

% \cite{Bell23} reported the detection of CH$_{4}$ detection at 6$\sigma$. This is further supported by the preference of our CH$_{4}$-rich models over H$_{2}$O-rich models in our transmission-only retrievals by 4$\sigma$ (Sect. \ref{sec:transmission_only}).

%This could be the case for exoplanets with equilibrium temperatures $T_{\rm eq}$ = 400 - 1000 K and masses $M < 1 \ M_{\rm Jup}$,  . 

In our fiducial retrieval, we assume equilibrium chemistry; however a high internal temperature combined with strong vertical mixing could drive the chemistry out of equilibrium. Methane is expected to be the dominant absorber in cold exoplanet atmospheres under chemical equilibrium \citep{Fortney20,Blain21}. Although our retrieved chemical abundances agree well with expectations for chemical equilibrium, the abundances of key absorbers (e.g., CH$_{4}$) may be sensitive to the temperature in the deep layers of the atmosphere \citep{Fortney20,OhnoFortney23}. Warm temperatures in these deep layers of the atmosphere can reduce the abundance of CH$_{4}$ by several orders of magnitude, regardless of the vertical diffusion coefficient, $K_{zz}$ \citep{Mukherjee25}. Thus, there is a degeneracy between $K_{\rm zz}$, $T_{\rm int}$ and C/O. If the deep atmosphere of WASP-80 b is heated (e.g., by Ohmic dissipation), the C/O ratio could be higher than the value we infer under equilibrium assumptions. Given the potential impact of disequilibrium chemistry, our future work will focus on incorporating it into our retrieval framework.

\subsection{Model uncertainties and caveats} \label{sec:caveats}

The standard deviation of the CMF posterior in the joint retrieval JR6 ($\Delta$CMF = 0.02) is smaller than the spacing of the interior model grid ($\Delta$CMF = 0.10). This could potentially lead to artificial precision if the assumption of linear interpolation between grid points -- considered by the Python function we use -- were inaccurate.  This could be the case in the regime of vanishingly small cores, where the planet radius could vary non-linearly with CMF. To assess this, we evaluated the accuracy of our linear interpolation method (\texttt{scipy.RegularGridInterpolator}) for the interpolated parameters: planetary radius and entropy at 1000 bar, and the $f_{S}' = f_{S} \times L$ parameter (see Sect. \ref{subsec:gastli_joint_mode}). Using models computed specifically for this test over the CMF = 0–0.10 range, we find that the maximum interpolation errors are $\sim$0.5\% for radius and $\sim$0.9\% for $f'_{S}$ -- well below the variations induced by $\Delta$CMF = 0.02, which corresponds to the uncertainty of the retrieved CMF. The entropy remains constant with CMF, and its interpolation error is negligible ($\sim10^{-6}$\%). These results confirm that the interpolation introduces errors significantly smaller than the posterior uncertainty, validating the robustness of our retrieval in the low-CMF regime.

% schwarzschild criterion
%If the gas opacity is sufficiently low, radiative layers and double-diffusive layers can increase the temperature in the deep layers of the atmosphere \citep{leconte12,radiative_layers_update}, thereby affecting the inference of the bulk composition. In future work, we will focus on implementing the Schwarzschild criterion in our modelling framework. Under this criterion, the temperature in the deep atmosphere ($P = 1-1000$ bar) would be calculated self-consistently rather than prescribed using the analytical P-T profile of \cite{Guillot2010}. Incorporating this approach into our modelling could potentially shift the bulk metal mass fraction of our fiducial joint retrieval, JR6, toward higher values. 

The formation of deep radiative and double-diffusive layers can impact the inferred bulk composition in gas giants. In the following, we assess the likelihood of exhibiting these in WASP-80 b. In the joint retrievals, we couple the \cite{Guillot2010} temperature–pressure profile to the interior model at 1000 bar. While this analytical prescription of the atmospheric thermal structure enables the formation of radiative layers, GASTLI assumes that the interior is fully convective. This means that our framework cannot form radiative layers at pressures $P>$ 1000 bar. The interior-only retrievals use self-consistent atmospheric models to compute the boundary condition for our forward models. These atmospheric models were obtained using petitCODE, which applies the Schwarzschild criterion to determine whether a layer is convective or radiative (see Sect. \ref{sec:gastli_normal_mode}). The pressure–temperature profiles of two of these atmospheric models are shown in Fig. \ref{fig:ptprofile_joint} for a solar composition and internal temperatures of $T_{\rm int}=$ 100 K and $T_{\rm int}=$ 500 K. The most favorable conditions for the formation of radiative layers are (1) low envelope metallicities and (2) low internal temperatures. From the two fiducial retrievals, the lowest atmospheric metallicity and internal temperature inferred for WASP-80 b are approximately $\sim 2 \times$ solar and $\sim$100 K, respectively. The petitCODE model at solar composition and $T_{\rm int}=$ 100 K (Fig. \ref{fig:ptprofile_joint}) already shows a convective zone extending from 100 bar to higher pressures, placing the radiative–convective boundary (RCB) of WASP-80 b at 100 bar. Increasing the internal temperature and atmospheric metallicity shifts the RCB upward (to lower pressures), as demonstrated by the model at $T_{\rm int}=$ 500 K. Moreover, the presence of clouds -- which are likely in WASP-80 b (see Sect. \ref{sec:disc_previous}) -- would further shift the RCB to lower pressures. Therefore, maintaining the RCB at pressures higher than 100 bar, as we do in our joint retrievals at 1000 bar, represents a conservative choice for the case of WASP-80 b and does not impact its inferred bulk composition.

If our framework were applied to a colder planet with a lower internal temperature or lower atmospheric metal content, fixing the RCB (i.e., the interior–atmosphere interface) at 1000 bar would indeed introduce a significant source of uncertainty in the inferred interior structure. For example, \cite{radiative_layers_update} find that Jupiter ($T_{\rm eq} =$ 122 K) could develop radiative layers at pressures between 1000 and 50 000 bar if its sodium (Na) and potassium (K) abundances were below $10^{-3} \times$ solar. As shown in their figure 8, the formation of radiative zones leads to colder interiors as the temperature profile becomes nearly isothermal compared to the dry adiabat. Therefore, implementing the Schwarzschild criterion in the interior structure model, GASTLI, will be within the scope of future work to extend its applicability to colder ($T_{\rm eq} < 800$ K) exoplanets.

Double diffusion -- also referred to as semiconvection -- may be relevant for WASP-80 b, as strong magnetic fields can inhibit large-scale, ideal convection, which is assumed in our interior model \citep{chabrier2007}. This effect could lead to imperfect mixing of heavy elements within the envelope. This process produces a superadiabatic temperature gradient, increasing the interior temperature relative to the ideal convective assumption \citep{leconte12}. Not treating double diffusion in our interior model may introduce a model uncertainty. Consequently, the statistical errors reported in Table \ref{tab:joint_intatm_summary} should be regarded as lower limits on the total uncertainty.

Finally, we discuss how the treatment of the He mass fraction in the envelope does not impact our results. GASTLI assumes a cosmogonic He mass fraction in the H-He mixture of the envelope of $Y = $ 0.275 \citep[see section 2.3.1 in][]{Acuna24_gastli_science_paper}. Given this assumption, the He mass fraction is treated consistently between the interior and the atmosphere in the free-chemistry joint retrieval. In the equilibrium-chemistry joint retrievals, the abundances of hydrogen and helium computed by easyCHEM are summed to calculate the total mass fraction of the H-He mixture. This approximation can lead to deviations at high atmospheric metallicities ($>250\times$ solar), where the He abundance tends to exceed the cosmogonic value. However, for WASP-80 b, the atmospheric metallicity retrieved in the fiducial equilibrium-chemistry joint retrieval (JR6) is 2-5$\times$ solar and the corresponding helium abundance remains close to $Y = 0.275$ in the H–He mixture. Therefore, this approximation does not affect our bulk composition estimates for WASP-80 b. We will implement a more self-consistent coupling between the helium abundance in the interior and the atmosphere in future work, particularly when applying our framework to exoplanets with highly metal-rich envelopes and observational constraints on helium.

Our model does not include conduction as a heat transport mechanism in the core. Given the available observables, the effect of conduction on the core's thermal structure would have a negligible impact on the total planetary radius. Nonetheless, conduction should be considered if magnetic field data are incorporated into an interior-atmosphere framework such as the one presented in this paper.

\subsection{Bulk composition and formation of WASP-80 b} \label{sec:formation_disc}

% (3) Planet formation of WASP-80 b and mass-bulk metallicity trend.

\begin{table*}[h]
\caption{\label{tab:summary_formation_params} Summary table of the atmospheric and bulk interior composition of WASP-80 b under the two fiducial scenarios.}
\centering
%\resizebox{\columnwidth}{!}{%
\begin{tabular}{lcc}
\hline
Parameter & Secular Cooling (JR6) & Additional heating (JR5$^{\ast}$) \\ \hline
Atmospheric metallicity, M/H [$\times$ solar] & $2.75^{+0.88}_{-0.56}$ & 10.00$^{+8.20}_{-4.75}$  \\
Atmospheric C/O & 0.12$^{+0.03}_{-0.02}$ & 0.24$\pm$0.09 \\ 
Core mass, $M_{\rm core} \ [M_{\oplus}]$ & 3.49$^{+3.49}_{-1.59}$ & 31.8$^{+21.3}_{-17.5}$  \\ 
Planetary bulk metal mass fraction, $Z_{\rm planet}$ & 0.12 $\pm 0.02$ &  0.28$\pm$0.11  \\ 
Planet-to-star bulk metal mass fraction ratio, $Z_{\rm planet}/Z_{\rm star}$ & $12.1^{+4.9}_{-5.1}$ & $30.3^{+15.5}_{-15.9}$  \\ \hline
\end{tabular}%
%}
\tablefoot{The first fiducial scenario is secular cooling, where no additional heating sources exist. The internal temperature is constrained by age and equilibrium chemistry is assumed in this scenario, whereas the second scenario corresponds to additional heating sources (i.e., Ohmic dissipation).}
\end{table*}

The formation history of a planet shapes both its atmospheric and bulk composition. Accordingly, four of the parameters obtained in this work -- atmospheric M/H, C/O ratio, core mass ($M_{\rm core}$) and bulk metal mass fraction ($Z_{\rm planet}$) -- allow us to explore potential formation and evolutionary pathways for WASP-80 b. To calculate the planet-to-star bulk metal mass fraction ratio, $Z_{\rm planet}/Z_{\rm star}$, we require the stellar bulk metal mass fraction. We estimate WASP-80's stellar bulk metal mass fraction using the approximation $Z_{\rm star} = 0.014 \times 10^{\rm [Fe/H]_{\star}}$ \citep[after][]{Thorngren16}, where [Fe/H]$_{\star} = -0.13^{+0.15}_{-0.17}$ \citep{Triaud15}.

We consider two scenarios: one in which heating mechanisms are absent, and another in which high temperatures in the deep atmosphere are induced by such mechanisms. In the former, thermal evolution is driven by secular cooling. Under this assumption, the planet's age constrains its internal temperature, which is the approach adopted in retrieval JR6. We adopt JR6 as our fiducial joint retrieval for this scenario because it consistently fits all available data - mass, age, transmission, and emission spectra. For the second scenario, the potential impact of heating sources on the inferred bulk and atmospheric composition is captured by our free-chemistry joint retrieval JR5$^{\ast}$. This retrieval is similar to JR6, but excludes the age constraint and does not assume chemical equilibrium. By omitting age as an observable, we remove the coupling between the planet's thermal cooling history and its internal temperature. This allows the model to explore thermal states that are hotter than those expected in the absence of heating mechanisms.

%WASP-80 b's bulk metal mass fraction \textbf{in the first scenario} is consistent with the trend derived by \cite{Thorngren16}, as predicted by the core accretion model (see Fig. \ref{fig:thorngren16_trend}, top panel). 

Table \ref{tab:summary_formation_params} summarizes the atmospheric and bulk composition of WASP-80 b under our two fiducial scenarios. In the first scenario (no heating, JR6), we obtain a core mass of $M_{\rm core, \ JR6}$ = 3.49$^{+3.49}_{-1.59} \ M_{\oplus}$. Core accretion typically produces cores with masses of $\sim$10 $M_{\oplus}$ \citep{Pollack96}, which is within the 3$\sigma$ upper limit of our estimate in the secular cooling scenario. Thus, the secular cooling scenario is marginally compatible with core accretion under highly favourable conditions. For example, \cite{Piso15} estimate that cores of masses as low as 3.5-5 $M_{\oplus}$ can be formed if core accretion occurs at long orbital distances (100 AU) and the disk opacity is low due to grain coagulation \citep[see also][]{Hubickyj05,Hori10}. Classic gravitational instability models can more easily reproduce such low-mass cores and even a core-less interior, as it is suggested by our first scenario within 2$\sigma$.

%star-to-planet metal mass fraction ratio of $(Z_{\rm planet}/Z_{\rm star})_{\rm JR6} = 12.1^{+4.9}_{-5.1}$ and a

% old JR6 discussion:
%This core mass could also be produced by gravitational instability, the second leading planet formation mechanism, under certain circumstances. Gravitational instability can form cores as massive as 10 $M_{\oplus}$ if the dust-to-gas ratio is high (1\%) and core formation begins in the early stages of protostellar evolution ($\sim$0.1 Myr). Spiral structures in the protoplanetary disk may further assist this process \citep{Rice25}.

In contrast, if a heating mechanism (e.g., Ohmic dissipation) raises the temperature in the deep envelope of WASP-80 b, its formation would be consistent with the core accretion paradigm within 1$\sigma$. Under our additional heating scenario, we estimate $M_{\rm core, \ JR5^{\ast}}$ = 31.8$^{+21.3}_{-17.5} \ M_{\oplus}$. This core mass is $\sim$10 times higher than that obtained in the first scenario with no additional heating sources, JR6. A core as massive as $\sim 30 \ M_{\oplus}$ can be produced in core accretion models. For example, pure pebble accretion models without planetesimal formation can reproduce a bulk metal mass fraction consistent with our JR5$^{\ast}$ estimate within 1$\sigma$ if the dust-to-gas ratio is high \citep[2\%, see figure 3 in][]{danti23}.

\begin{figure}[h]
   \centering
   \includegraphics[width=\columnwidth]{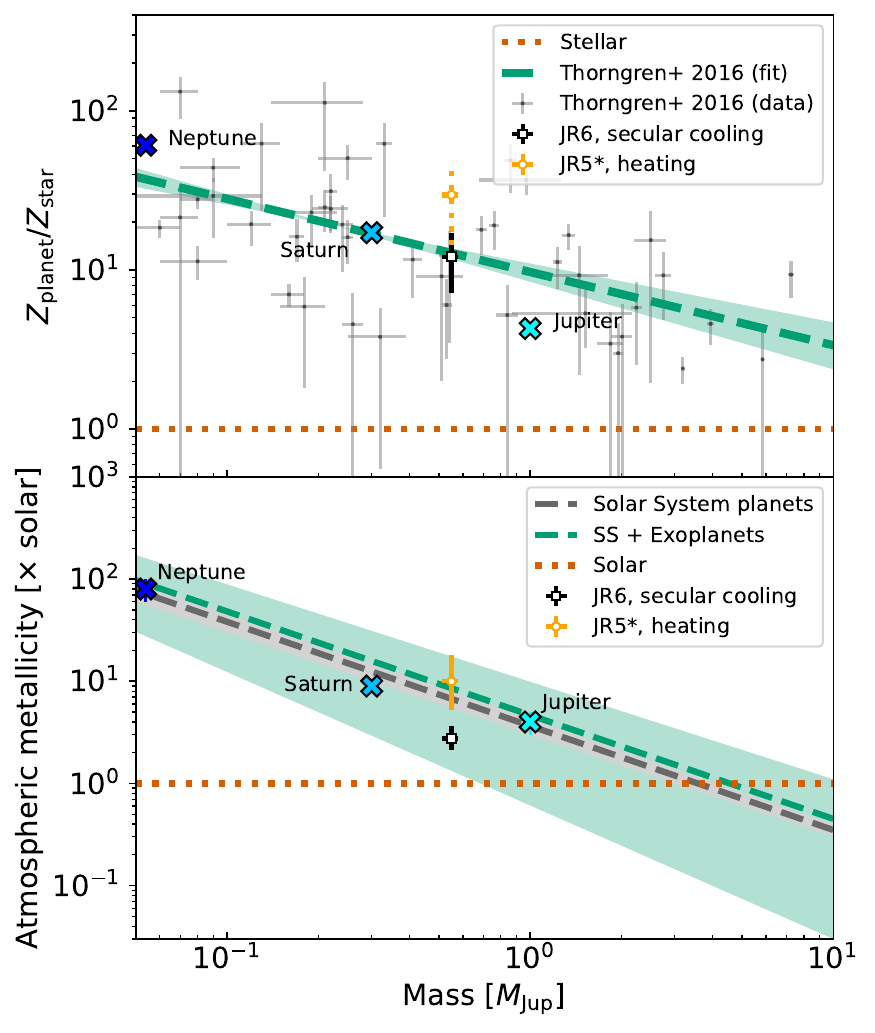}
      \caption{Planet mass-bulk metal mass fraction diagram (top panel) and mass-atmospheric metallicity diagram (bottom panel) for WASP-80 b. We show the estimates of WASP-80 b's for our two fiducial retrievals: JR6 (black; consistent with all data and no heating mechanism) and its heating counterpart, JR5$^{\ast}$ (orange; free chemistry and no age constraint). The bulk and atmospheric metallicity trend fits are adopted from \cite{Thorngren16} and \cite{wakeford18}, respectively. We compare with estimates for Jupiter \citep{Howard23}, Saturn \citep{Militzer19,MV23} and Neptune \citep{Helled11,Karkosha}.}
         \label{fig:thorngren16_trend}
\end{figure}

Exoplanets continue to accrete solids and gas into their envelope after reaching the pebble isolation mass \citep{Pollack96,Turrini15,Emsen21}. The atmospheric metallicity and C/O ratio therefore contain information about the composition of the accreted gas, its location, and the dominant accretion mechanism (pebbles vs planetesimals). In our first scenario JR6, we retrieve an atmospheric metallicity of M/H = 2.75$^{+0.88}_{-0.56} \times$ solar and a C/O ratio of 0.12$^{+0.03}_{-0.02}$. Although the C/O ratio retrieved in JR5$^{\ast}$ -- which represents a scenario with additional heating and no age constraint -- is slightly higher, it remains sub-solar (C/O$_{\rm JR5^{\ast}}$ = 0.24$\pm$0.09). A sub-solar C/O ratio suggests that WASP-80 b accreted oxygen-rich material. This material could be H$_{2}$O ice accreted outside the water snowline via planetesimal accretion. Alternatively, WASP-80 b may have accreted gas enriched in water vapor inside the snowline, especially in our heating scenario, where the atmospheric metallicity is as high as 10-20$\times$ solar. Pebbles outside the water ice line can drift inwards and evaporate as they cross this line. Inward-drifting pebbles from beyond the snowline can evaporate upon crossing it, enriching the inner disk gas with water vapor. Low C/O ratios are particularly favored if inward migration is rapid, a condition facilitated by high-viscosity disks \citep{Schneider21}.

%%%% STOP HERE %%%%

%\textcolor{RoyalBlue}{Update this paragraph:} Our analysis suggests that WASP-80 b has a sub-solar atmospheric metallicity, which is unexpected for a planet that formed via core accretion as these planets tend to have super-solar envelopes because they accrete solids during and after envelope formation. A sub-solar [M/H] can indicate that WASP-80 b formed via pebble accretion. In this scenario, the planet rapidly builds a core but halts solid accretion when the pebble isolation mass is reached \citep{Lambrechts14,Bitsch18,Ataiee18}. If this happens early, before metals are mixed into the atmosphere, it may lead to a low atmospheric metallicity. Furthermore, planetesimal accretion is a source of metal enrichment of the envelope once the pebble isolation mass is reached \citep{Pollack96,Turrini15,Emsen21}. Given the sub-solar metallicity of WASP-80 b's atmosphere, it could have formed in an environment where planetesimal accretion was inefficient or absent during gas accretion. Shortly after accretion, planets may migrate inwards to their current orbit. Tidal circularization of high-eccentricity orbits often prevents the planet from passing through dense gas regions that would enable further accretion \citep{Petrovich}, which would also contribute to explaining WASP-80 b's sub-solar atmospheric metallicity. WASP-80 b probably accreted its atmosphere from unprocessed, solar-like gas inside the CO and H$_{2}$O snowlines, where the gas C/O is approximately the stellar C/O \citep{Oberg11,Booth19}, producing an atmosphere with a solar C/O. 

Finally, in Fig. \ref{fig:thorngren16_trend} we compare the planet-to-star bulk metal mass fraction ratio and atmospheric metalliticy for our two fiducial scenarios with the population trends derived by \cite{Thorngren16} and \cite{wakeford18}. Our two scenarios are consistent within 1$\sigma$ with the atmospheric metallicity of the general exoplanet population. The mean bulk metal content agrees extremely well with the population trend in our secular cooling scenario, whereas the heating scenario estimate agrees within its 1$\sigma$ uncertainty. \cite{Thorngren16} discuss that this mass-bulk metal mass fraction trend is a result of core accretion as the dominant formation mechanism in the exoplanet population. However, recent work by \cite{chachan25} has revisited this trend with a larger sample (150 planets), and found that the trend flattens at high masses ($> 1 \ M_{\rm Jup}$. This indicates that gravitational instability may play a more prominent role in the formation of the exoplanet population than previously suggested, or that significant heavy-element enrichment continues after the runaway accretion stage via accretion of pebbles and planetesimals. Our two interior scenarios are compatible with this late heavy-element enrichment, while our secular cooling scenario is consistent with gravitational instability.

\section{Conclusions} \label{sec:conclusions}

In this work, we carried out a suite of interior, atmosphere and joint interior-atmosphere retrievals for the warm gas giant WASP-80 b. Joint retrievals combine a coupled, self-consistent interior-atmosphere model with Bayesian inference, in which the log-likelihood incorporates information from both the spectrum and bulk density observables (mass, radius, age). The unique combination of transmission and emission spectra across a broad wavelength range, together with precise mass and age, allowed us to constrain the planet’s bulk metal content, atmospheric composition and thermal structure with unprecedented precision.

Based on atmospheric data alone, we find that transmission-only retrievals favor sub-solar to solar atmospheric metallicities with a C/O ratio consistent with solar, under the assumptions of chemical equilibrium and a wavelength-dependent cloud model. In contrast, free-chemistry retrievals and those employing simpler (e.g., grey cloud) models yield super-solar metallicities and low C/O ratios, consistent with previous work \citep{Wong22,Bell23}. We demonstrated that cloud treatment significantly affects the inferred atmospheric composition -- particularly the metallicity and C/O ratio -- and that equilibrium chemistry retrievals with wavelength-dependent clouds offer the best fit to the data based on Bayesian evidence. Retrievals based on emission data alone also support a super-solar metallicity and a sub-solar C/O ratio. Additionally, they constrain the thermal structure at pressures $P = 1-10^{-3}$ bar to temperatures $T = 700-1000$ K at 1$\sigma$.

Traditional interior-only retrievals treat mass, radius, age, and atmospheric metallicity as observables. For WASP-80 b, we find that interior retrievals that are consistent with the planet's age in the absence of additional heating sources favor a low envelope metallicity (M/H $<10 \times$ solar at 1$\sigma$). The higher metallicities suggested by the free transmission retrievals require extreme, high internal temperatures that are inconsistent with the age constraint. This implies that the highly enriched atmospheric metallicities inferred in previous atmospheric studies -- and in our transmission free-chemistry retrievals -- are only compatible with WASP-80 b’s bulk density if there are additional heating mechanisms (e.g., Ohmic dissipation) and/or an interior that is not fully dominated by convection (non-adiabatic).

%Taking into account the atmosphere and interior information in 

We thus adopt two fiducial joint interior-atmosphere retrievals that encapsulate different scenarios. The retrieval JR6 takes into account all available data -- mass, age, and transmission and emission spectra -- and represents a scenario where the planet's internal temperature is driven by secular cooling without additional heating sources. Under this scenario, WASP-80 b has an atmospheric metallicity of M/H = $2.75^{+0.88}_{-0.56} \times$ solar, a sub-solar C/O = 0.12$^{+0.03}_{-0.02}$, a bulk metal mass fraction $Z_{\rm planet}=0.12 \pm 0.02$ and a core mass of $M_{\rm core}$ = 3.49$^{+3.49}_{-1.59} \ M_{\oplus}$. On the other hand, the retrieval JR5$^\ast$ represents a second scenario where additional heating sources warm the deep envelope by decoupling the internal temperature from secular cooling and the assumption of equilibrium chemistry. In this inflated scenario, WASP-80 b has an atmospheric metallicity of M/H = 10.00$^{+8.20}_{-4.75} \ \times$ solar, a C/O = 0.24$\pm$0.09, a bulk metal mass fraction ratio $Z_{\rm planet}/Z_{\rm star}=30.3^{+15.5}_{-15.9}$ and a core mass $M_{\rm core}$ = 31.8$^{+21.3}_{-17.5} \ M_{\oplus}$. In addition, our findings are summarized as follows:

\begin{itemize}

\item We demonstrate that joint retrievals reveal degeneracies between envelope composition, chemistry and thermal state. The comparison between the joint retrievals JR1 and JR1$^{\ast}$ indicates the degeneracy between composition and chemistry, where WASP-80 b could have a (1) super-solar envelope and very low C/O ratio or (2) a sub-solar envelope in chemical equilibrium and C/O$\sim$0.40. Similarly, the contrast between retrievals JR5 and JR5$^{\ast}$ show the degeneracy between chemistry and thermal state (see Fig. \ref{fig:ptprofile_joint}, right panel), where equilibrium chemistry requires higher internal temperatures to fit the spectral data and bulk density simultaneously.

%reduce the tension between free-chemistry atmosphere-only and interior-only analyses, yielding atmospheric metallicities and cloud locations that are compatible with both spectral data and bulk density constraints. Thus, joint retrievals can resolve biases in atmospheric properties arising from systematic errors in the spectra or atmosphere models.

\item For planets older than 1 Gyr, the inclusion of age in interior-only retrievals does not significantly improve the estimate of the bulk metal mass fraction. However, in joint retrievals that include both emission and transmission spectra, including age as an observable reduces the uncertainties in the inferred CMF and bulk metal mass fraction by a factor of two. 

% The combination of the observational constraints -- age, transmission and emission spectra -- reduces the degeneracies between composition and thermal state. 

\item The addition of the panchromatic emission spectrum (1-12 $\mu$m) to the joint retrievals allows us to constrain the chemical abundances and the temperature at $P = 1-10^{-3}$ bar with high precision. This information, when combined with equilibrium chemistry models, allows us to break the degeneracy between temperature and envelope metal content. 

\item If the assumption of equilibrium chemistry is relaxed by adopting free chemistry, the uncertainty in the bulk metal mass fraction increases from $\Delta Z_{\rm planet, \ JR5} = 0.02$ (16\% precision) to $\Delta Z_{\rm planet, \ JR5^{\ast}} = 0.11$ (40\% precision). Thus, the degeneracy between atmospheric chemistry (equilibrium vs free) and internal temperature can increase the uncertainty in derived bulk metal mass fraction and core mass if the age is unconstrained or the planet is exposed to additional heating sources (i.e., tidal heating, Ohmic dissipation).

%only if the wavelength coverage is wide . Joint retrievals with limited emission coverage (1-4 $\mu$m) did not break this degeneracy, obtaining worse constraints than joint transmission retrievals in both bulk composition properties (CMF, $Z_{\rm planet}$) and atmospheric metallicity and C/O. 

%\item Joint retrievals with transmission and emission spectra improve uncertainties in bulk metal mass fraction by a factor of three in comparison to traditional interior-only retrievals, even if age is not available at all. 

%Joint retrievals that consider the bulk properties and the emission spectrum - without transmission - yield the largest uncertainties in inferred bulk metal mass fraction and atmospheric metallicity in the suite. This is caused by the temperature-$Z_{\rm env}$ degeneracy between 1 and 1000 bar. In addition, emission is only sensitive to CH$_{4}$, obtaining worse constraints in atmospheric composition than transmission.

%\item Joint retrievals that include transmission with bulk properties - without emission - seem to have the tightest constraints in bulk metal mass fraction and CMF. However, including emission spectra yields more accurate estimates as it determines the thermal structure at $P<1$ bar.

\end{itemize}

% Importantly, we show that combining transmission spectra with bulk observables offers the most precise and consistent estimate of the planet’s bulk metal mass fraction. Joint retrievals using emission spectra alone yield wider uncertainties due to degeneracies in the temperature–metallicity space at pressures >1 bar. Moreover, 

We also caution against biases in joint retrievals. Specifically, we advise against assuming clear-atmospheric forward models see Sect. \ref{sec:cloud_wilkinson}) and using weighting factors in the likelihood function to give more importance to bulk properties (see Sect. \ref{sec:likelihood_wilkinson}). The former eliminates the degeneracy between cloud location and envelope metal content while the latter can spuriously reduce the uncertainties in mass, radius, and age. Both effects lead to an underestimation of the uncertainties in the bulk metal mass fraction and CMF.

Finally, we explore possible caveats in our model assumptions about the interior and atmosphere. For the interior model, we assume that the deep interior follows an adiabat. Although WASP-80 b is unlikely to have compositional gradients or tidal heating, Ohmic dissipation may further heat its interior. This mechanism and its implications for the interior and atmosphere of sub-Saturn exoplanets at intermediate equilibrium temperatures should be explored in future work. Similarly, in our fiducial joint retrievals, we adopt either equilibrium or free chemistry. These two assumptions fit the spectra well, but we note a degeneracy between atmospheric composition and the extent of vertical mixing, $K_{\rm zz}$. In future work, we will incorporate the effects of vertical mixing with self-consistent chemistry grids, as well as constraints from reflected light and Bond albedo.

%Further work focused on intermediate equilibrium temperatures and masses is needed to determine whether Ohmic dissipation could be a significant source of heating in WASP-80 b. 

We find that WASP-80 b has an atmospheric and bulk composition consistent with the general exoplanet population within 1$\sigma$ in our two fiducial scenarios. The very low core mass derived in the first scenario (secular cooling) is consistent with gravitational instability, and could marginally be produced by core accretion under very favourable conditions. In contrast, the second scenario (additional heating) agrees very well with the classical core accretion paradigm. Super-solar atmospheric metallicities -- $\sim2$ and $\sim10\times$ solar in the secular and additional heating scenario, respectively -- and sub-solar C/O ratios suggests that WASP-80 b not only accreted solids during core formation but also continued to do so afterward. This envelope enrichment is likely produced by planetesimal-dominated accretion beyond the water snowline or pure pebble accretion within the water ice line.

%Our study highlights the power of joint interior-atmosphere retrievals to infer robust planetary bulk and atmospheric compositions, particularly when multi-geometry spectra and age estimates are available. 

Our study highlights that for WASP-80 b, joint interior-atmosphere retrievals have degeneracies between atmospheric and interior processes, and robust inferences of atmospheric metallicity, C/O ratio and bulk metal mass fraction require precise mass, radius, age and atmospheric spectra combined with this approach. Future observations combining transmission, emission, and reflected light spectroscopy -- especially with JWST and upcoming missions such as Ariel -- will enable comprehensive interior-atmosphere modeling for a broader sample of exoplanets. This will be essential for understanding the connection between planet formation and composition at a population level.

%These refined compositional estimates offer a coherent picture of a modest core ($M_{\rm core}$ = $3.49^{+3.49}_{-1.59} \ M_{\oplus}$) and a solar to super-solar envelope. 

% Gather main conclusions and takeaways of the paper: (1) joint interior-atmosphere retrievals improve precision of bulk metallicity, and can resolve the tension between independent interior and atmosphere retrievals; 

% (2) clouds need to be taken into account, otherwise the estimates on CMF and [M/H] are biased. 

% Discuss a bit the future work, such as planets with disequilibrium chemistry or sub-Neptunes.

\begin{acknowledgements}
      % Acknowledgements go here
      L. A. A. thanks Bertram Bitsch, Eva-Maria Ahrer and Shreyas Vissapragada for insightful discussions on planet formation, atmospheric data analysis and exoplanet magnetic fields during the scope of this work.
\end{acknowledgements}


\bibliographystyle{aa}           % style aa.bst 
\bibliography{references}      %% example.bib = bibtex 

\begin{appendix} 

% Second appendix
\section{Forward models for traditional interior retrievals}

\begin{figure}[h]
   \centering
   \includegraphics[width=\columnwidth]{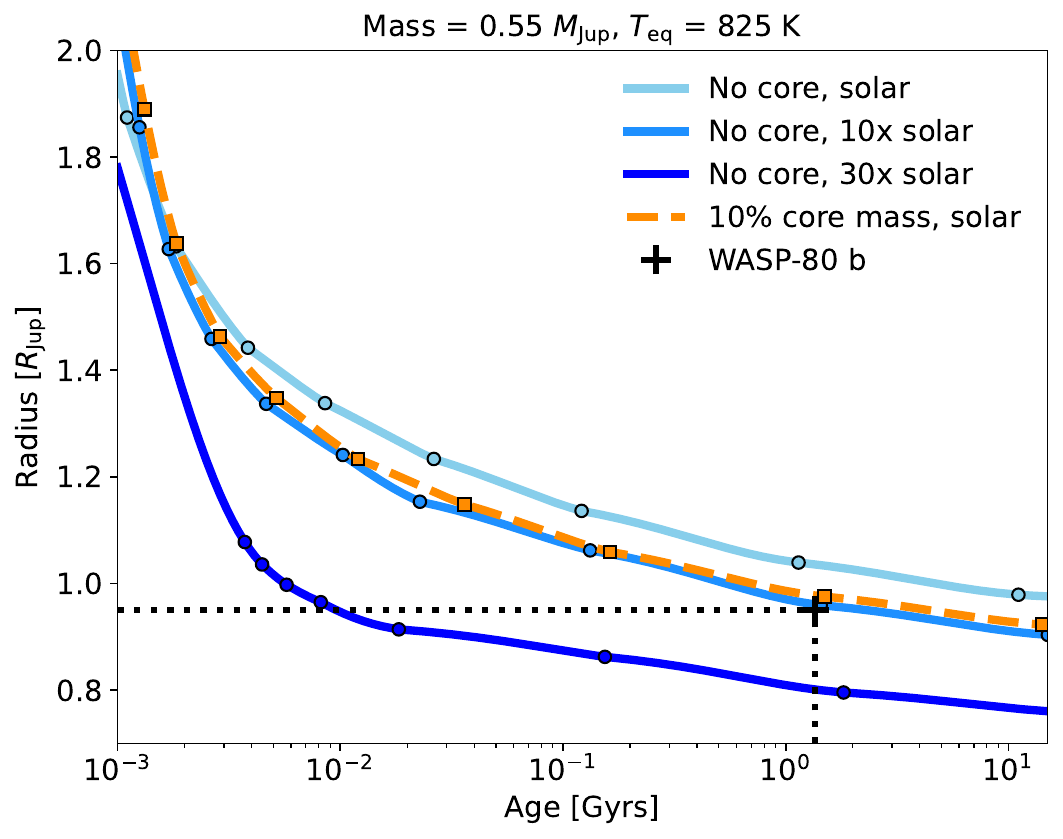}
      \caption{Radius-age relations of WASP-80 b. These were obtained from the grid of GASTLI models described in Sect. \ref{subsec:gastli_normal_mode}}
         \label{fig:interior_forward_nocore}
\end{figure}

Fig. \ref{fig:interior_forward_nocore} shows four thermal evolution curves -- radius-age relations --evaluated at WASP-80 b's mean mass and equilibrium temperature. These thermal evolution curves constitute the forward model in the traditional interior retrievals (Sect. \ref{sec:interior_only_retrievals}).  These forward models suggest that the mass, radius and age of WASP-80 b are compatible with a 10$\times$ solar composition envelope with no core or a 10\% core mass with a solar envelope. Compositions greater than 30$\times$ are ruled out by more than 4$\sigma.$

% APPENDIX 0.0

\section{Interior-atmosphere coupling for joint retrievals.}
\label{sec:coupling_joint}

\begin{figure}[h]
   \centering
   \includegraphics[width=\columnwidth]{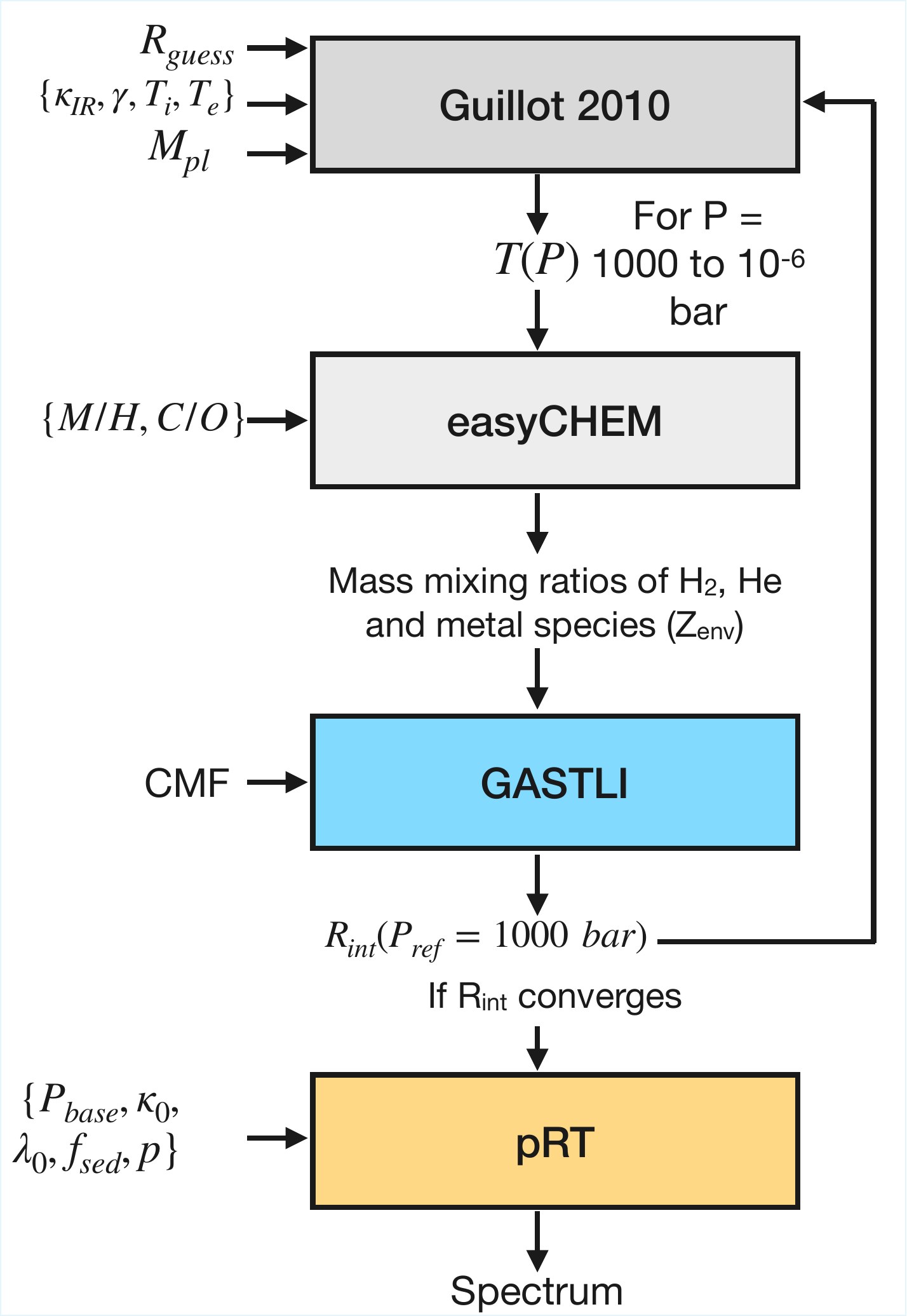}
      \caption{Modified coupling algorithm. $M_{\rm pl}$ is the planetary mass, while $R_{\rm int}$ is the radius from the planetary center up to the interior-atmosphere interface, located at $P$ = 1000 bar. $R_{\rm guess}$ is the initial guess value for the radius, $R_{\rm int}$. The remaining 10 parameters correspond to the P-T analytical model (Guillot 2010; $\kappa_{\rm IR}$, $\gamma$, $T_{\rm i}$, $T_{\rm e}$), the chemical equilibrium grid (\textit{easyChem}; M/H, C/O), GASTLI's interior structure module (CMF); and \textit{pRT} cloud opacity parametrization ($P_{\rm base}$, $\kappa_{0}$, $\lambda_{0}$, $f_{\rm sed}$, $p$).
      }
         \label{fig:joint_coupling_algorithm}
\end{figure}

% New coupling method between interior model and petitradtrans. 

In Sect. \ref{subsec:gastli_normal_mode} we recapitulate our approach to couple self-consistently GASTLI's interior model to a grid of 1D atmospheric models consisting of temperature (P-T) and metal mass fraction (P-$Z_{\rm env}$) profiles. Nonetheless, this coupling algorithm requires to be adapted when coupling the new interior model grid dedicated to joint retrievals (see Sect. \ref{subsec:gastli_joint_mode}) to \verb|pRT| due to the new set of input parameters.

% If enough room, add a figure with a diagram of the algorithm for clarity. 

Fig. \ref{fig:joint_coupling_algorithm} illustrates the modified coupling algorithm used in the joint atmosphere-interior retrievals. The planet mass $M_{\rm pl}$ and the other 10 free parameters (see caption) are constant across one instance of the coupled model. The internal planet radius (center to 1000 bar), $R_{\rm int}$ is updated after the grid of GASTLI's interior structure models is interpolated. The condition for radius convergence is similar to that presented in \cite{Acuna21}: the difference between successive radius calculations is lower than a specified tolerance, $\Delta_{\rm tol} = 10^{-3} \times R_{\rm int}$. GASTLI's grid not only requires the envelope metal mass fraction estimated by the chemical equilibrium grid, $Z_{\rm env}$ but also the temperature at 1000 bar, $T_{\rm surf}$, computed in the first step by the analytical P-T profile. Thus, in the last step where the transmission spectrum is computed by \verb|pRT|, the reference radius is equal to the converged $R_{\rm int}$ value, being specified at a reference pressure of 1000 bar.

\begin{figure}[h]
   \centering
   \includegraphics[width=\columnwidth]{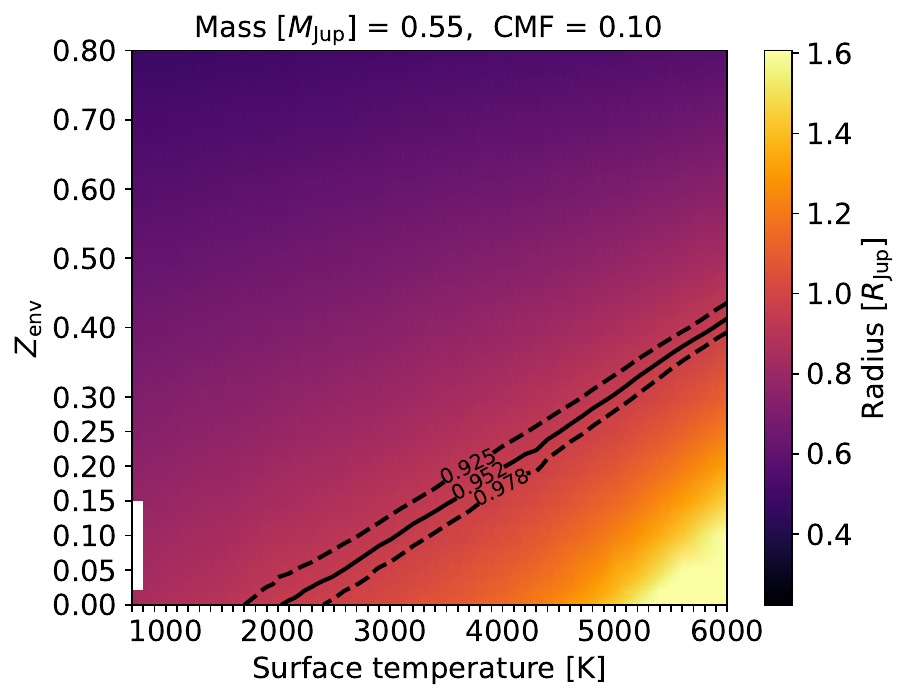}
      \caption{Interior planet radius ($R_{\rm int}$, defined at a pressure of 1000 bar) as a function of envelope metal mass fraction $Z_{\rm env}$ and surface temperature ($T_{\rm surf}$, also defined at 1000 bar). These models assume a constant mass $M = 0.55 \ M_{\rm Jup}$ and CMF = 0.10. Solid and dashed black lines indicate the mean and 1$\sigma$ range of WASP-80 b's measured radius (Table \ref{tab:wasp80b_all_obs_data_summary}).}
         \label{fig:joint_grid_example}
\end{figure}

Fig. \ref{fig:joint_grid_example} shows a slice of the multidimensional grid evaluated at the mean mass of WASP-80 b and a CMF = 0.10. The radius is a smooth function of surface temperature and envelope metal mass fraction, demonstrating that our grid is sufficiently finely sampled. This fine sampling allows us to interpolate the radius and entropy with relative errors of $<1$\% with respect to GASTLI computed models, which is well under the observational errors of $3$\%. WASP-80 b's measured radius (solid black line) is slightly larger than the interior radius ($R_{\rm int}$) as the atmosphere contributes to the total radius from 1000 bar to $\sim$20 mbar. Three models at low temperatures (T = 700 K) and low envelope metal mass fractions did not converge (white). This does not impact our grid interpolation and retrieval framework, as this is well below the equilibrium temperature of WASP-80 b (825 K).

% First appendix
%\newpage

\section{Best-fit models}

% Figure (2) plot of the nominal fit + spectrum. You can mention the chi square in the caption, just to show that it is a good fit.

\begin{figure*}[h]
   \centering
   \includegraphics[width=0.9\textwidth]{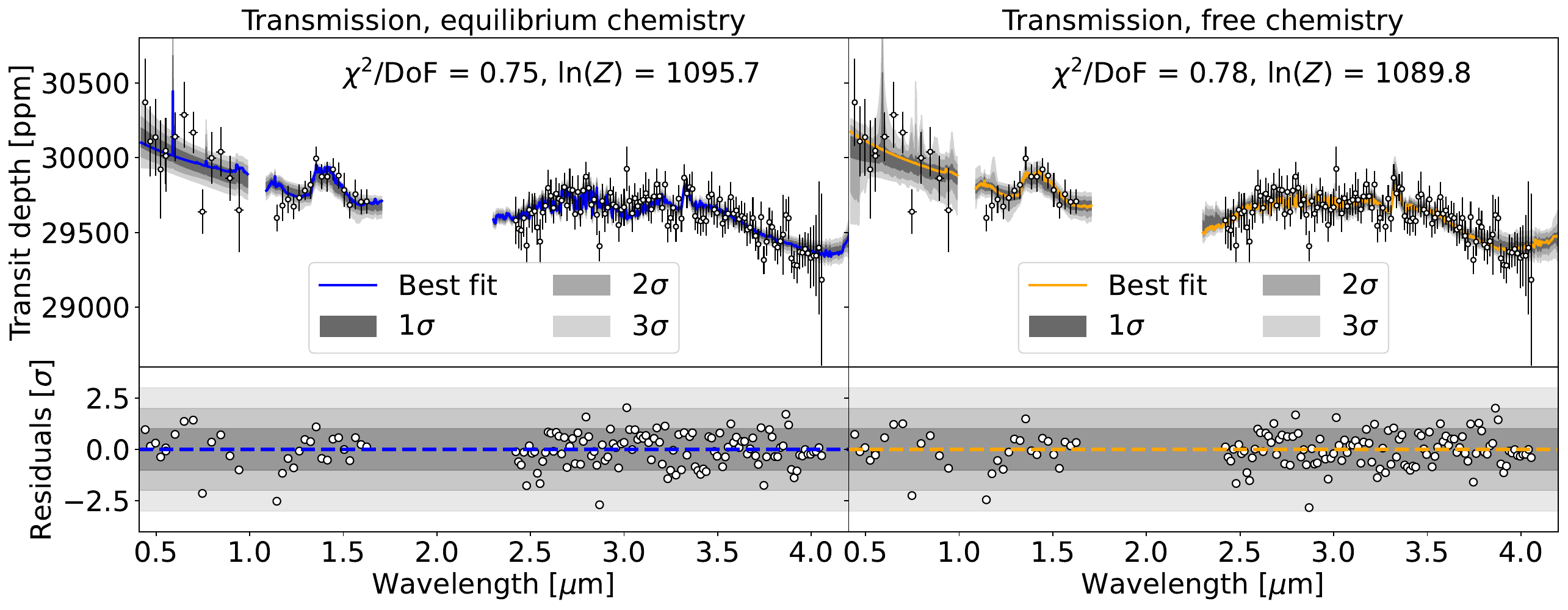}
      \caption{Best-fit model for our transmission spectrum retrievals of WASP-80 b with equilibrium (left panel) and free (right panel) chemistry.}
         \label{fig:transmission_retrieval_bestfit}
\end{figure*}

\begin{figure*}[h]
   \centering
   \includegraphics[width=0.9\textwidth]{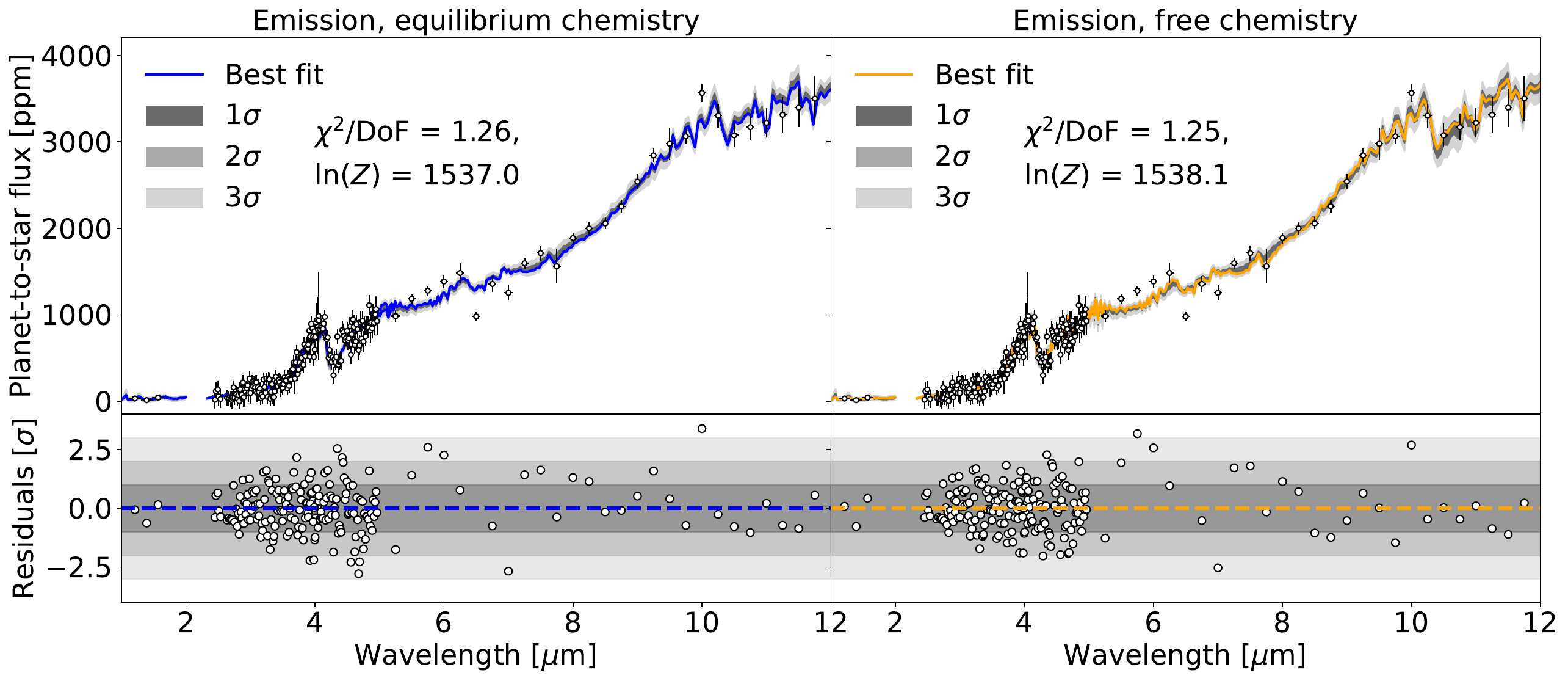}
      \caption{Best-fit model for our emission spectrum retrievals of WASP-80 b with equilibrium (left panel) and free (right panel) chemistry.}
         \label{fig:emission_retrieval_bestfit}
\end{figure*}

\begin{figure*}[h]
   \centering
   \includegraphics[width=0.9\textwidth]{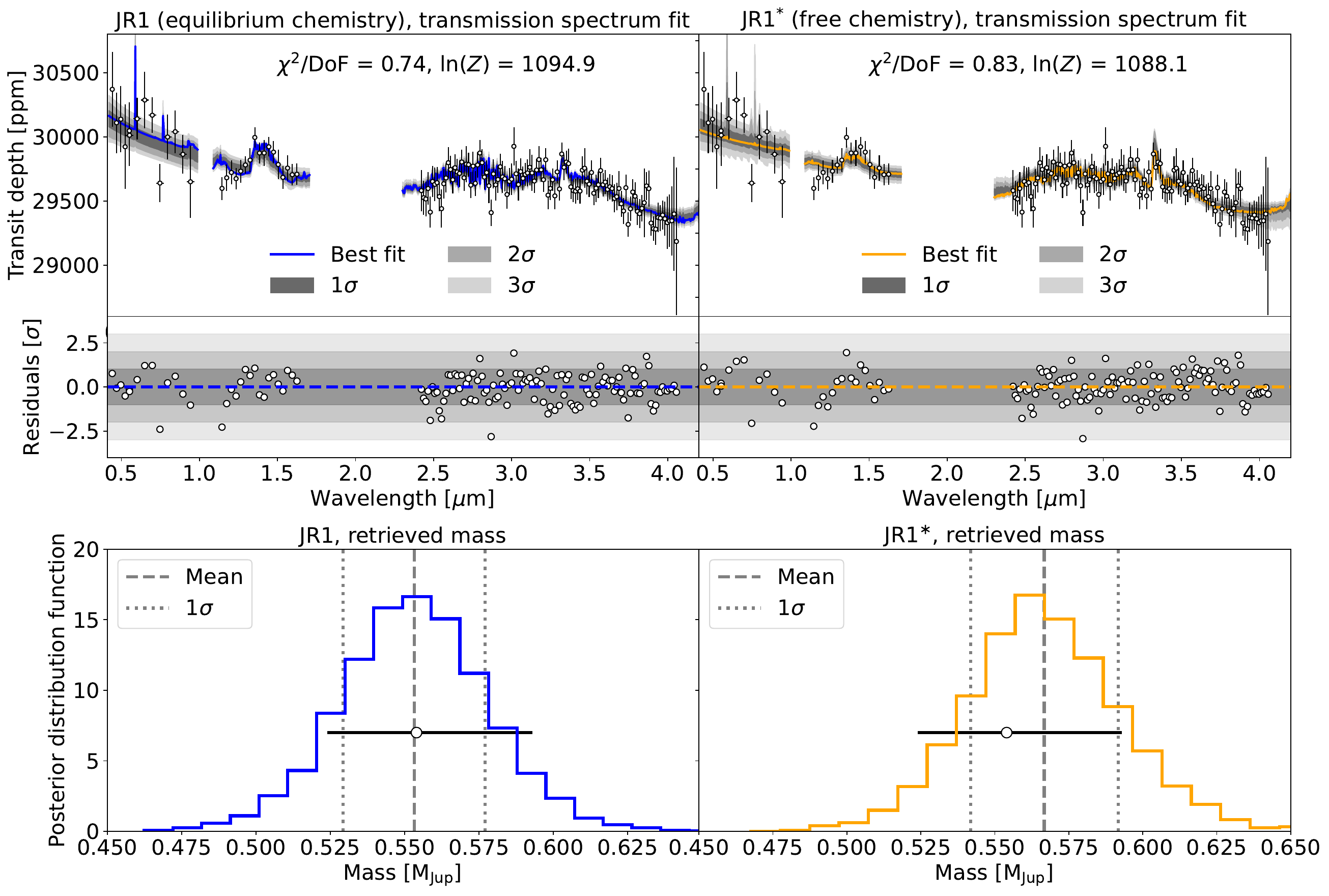}
      \caption{Top row: Best-fit models for our joint retrievals of WASP-80 b JR1 (equilibrium chemistry; left panel) and JR1$^{\ast}$ (free chemistry; right panel). Bottom row: Posterior distribution function (PDF) of the retrieved masses for JR1 and JR1$^{\ast}$. Dashed and dotted lines indicate the mean and standard deviations of the PDFs. White circles indicate the observed data with their respective uncertainties in all four panels.}
         \label{fig:jointMASS_retrieval_bestfit}
\end{figure*}

\begin{figure*}[h]
   \centering
   \includegraphics[width=0.9\textwidth]{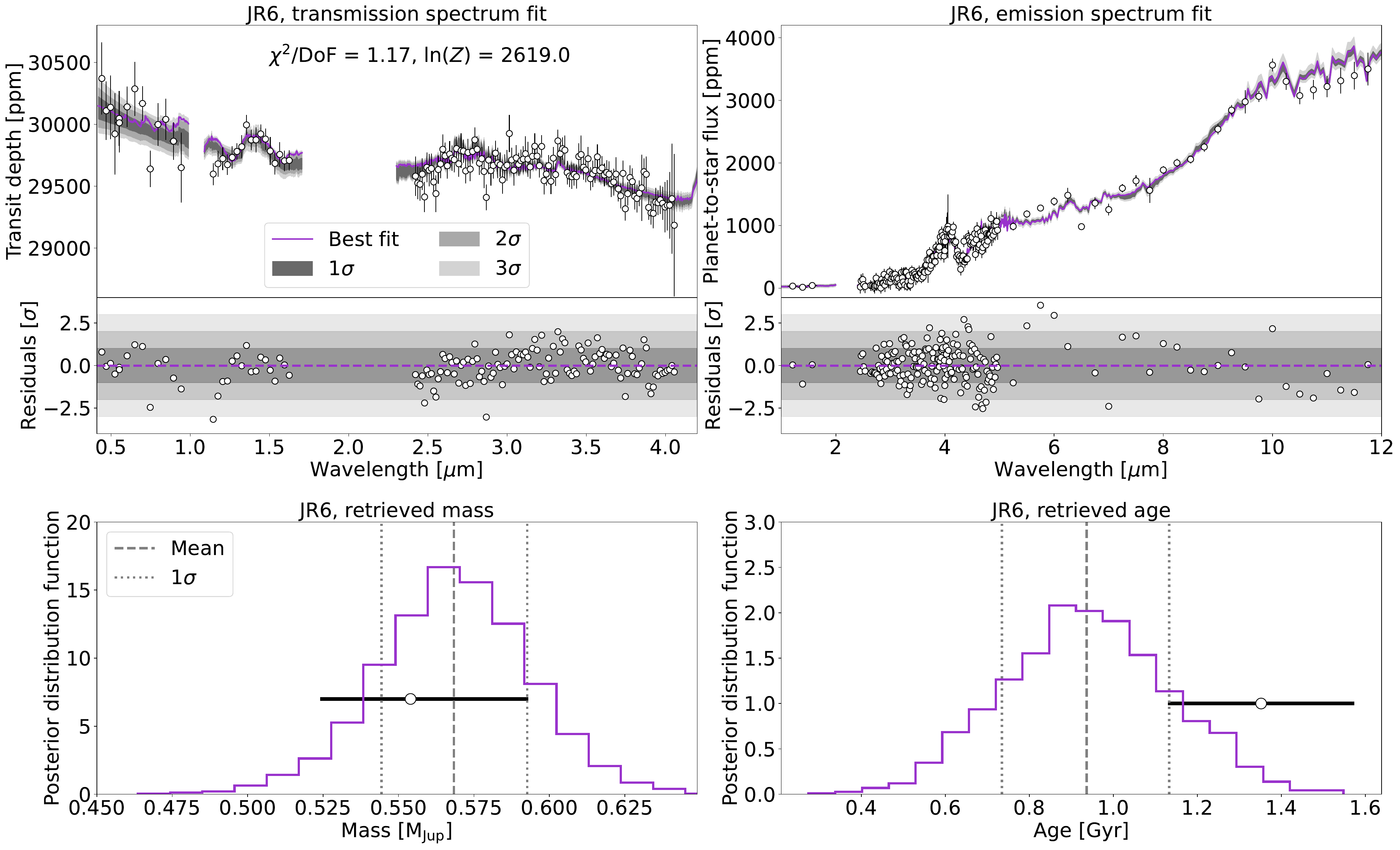}
      \caption{Best-fit model for our joint retrieval JR6.}
         \label{fig:jointALL_retrieval_bestfit}
\end{figure*}

% Third appendix
%\section{Thermal structure in joint retrievals}

\end{appendix} 

\end{document}